\documentclass[showpacs,aps,twocolumn,prb,groupedaddress]{revtex4-1}

\usepackage{graphicx}
\usepackage{color}
\usepackage{amsmath,verbatim}
\usepackage{amssymb,ulem}

\renewcommand{\emph}[1]{\textit{#1}}

\newcommand{\vek}[1]{\underline{#1}}
\newcommand{\vk}{\underline{k}}
\newcommand {\om}  {\ensuremath{\omega}}
\renewcommand{\Im}{\ensuremath{\mathrm{Im}}}
\newcommand{\si}{\sigma}
\newcommand{\fd}{{\hat f}^{\dagger}}
\newcommand{\f}{\ensuremath{\hat f^{\phantom{\dagger}}}}
\newcommand{\eps}{\epsilon}
\newcommand {\cd}  {\ensuremath{\hat c^{\dagger}}}
\renewcommand {\c}  {\ensuremath{\hat c^{\phantom{\dagger}}}}
\newcommand {\ull}[1] {\ensuremath{\underline{\underline{#1}}}}

\newcommand {\partition} {\ensuremath{\mathcal{Z}}}

\newcommand{\m}[1]{\mathrm{#1}}

\begin{document}

\title{Spectral properties of the two-impurity Anderson model with varying distance and
various interactions}

\author{Torben Jabben and Norbert Grewe}
\affiliation{Institut f\"ur Festk\"orperphysik, Technische Universit\"at Darmstadt, 
  Hochschulstr.\ 8,  64289 Darmstadt, Germany}
\author{Sebastian Schmitt}
\email[Email: ]{sebastian.schmitt@tu-dortmund.de}
\affiliation{Lehrstuhl f\"ur Theoretische Physik II, Technische Universit\"at Dortmund,
  Otto-Hahn-Str.\ 4, 44221 Dortmund,  Germany}

\date{\today}

\begin{abstract}

  We present a novel approximation for the treatment of two 
  interacting magnetic impurities immersed into a noninteracting
  metallic host. The scheme is based on direct perturbation theory 
  with respect to the hybridization between the impurity and band 
  electrons. This two-impurity enhanced noncrossing 
  approximation can fully incorporate the indirect
  interactions between the impurities which are mediated by the conduction electrons
  as well as additional
  arbitrary  direct interaction matrix elements. We qualify the
  approximation by investigating the uncoupled case 
  and conclude the two-impurity approximation to be 
  equally accurate as its single impurity counterpart.
  The physical properties of the two-impurity Anderson model
  (TIAM) are first investigated in some limiting cases.
  In a regime, where each of the uncoupled two impurities would exhibit
  a pronounced  Kondo effect, we ignore the indirect coupling via the conduction 
  band and only incorporate direct interactions. 
  For a ferromagnetic direct exchange coupling, the system displays a behavior
  similar to a spin-one Kondo effect, while an  antiferromagnetic 
  coupling competes with the Kondo effect and produces a pseudogap 
  in the many-body Kondo resonance of the single-particle spectral function.
  Interestingly, a direct one-particle hopping also produces a
  pseudogap but additionally pronounced side-peaks emerge. This gap is characteristically
  different from the case with antiferromagnetic coupling, 
  since it emerges as a consequence of distinct Kondo effects 
  for the bonding and anti-bonding orbital, i.e.\ it reflects a
  splitting of even and odd parity states.    
  For the general case of only indirect coupling via the conduction band  
  the results show signatures of all the previously discussed limiting cases
  as function  of the impurity-impurity distance. Oscillatory behavior 
  in physical quantities is to be  expected due to the generated 
  Ruderman-Kittel-Kasuya-Yosida (RKKY)-interaction.
  We are led to the conclusion that the well known Doniach-scenario captures 
  essential aspects this model,   but the details  especially at small distances 
  are more complicated.
  
\end{abstract}





\maketitle

\section{Introduction}
\label{sec:intro}

The investigation of strong correlation effects plays a major 
role in contemporary condensed matter research\cite{colemanUnfinishedFrontier03}
as, for example, in the field of high-temperature cuprate
superconductors,\cite{plakidaHighTCBook10} heavy-fermion
systems\cite{greweSteglichHF91,*colemanHeavyFermionMag07,*steglich:HFQCP06,*siQCHF10}  
or frustrated magnetism.\cite{lacroixFrust2010}
Due to the complexity of the phenomena and the theoretical models there
are still fundamental questions to be addressed, despite 
long and intensive research in these fields. 

Quite generally, the reason for this can be found in the presence of 
competing physical tendencies, whose characteristic energy scales are 
close together.\cite{ElbioDagotto07082005}
Then, a very rich variety of physical scenarios 
results since subtle differences in the 
external conditions or small parameter changes can tip the 
balance from one  dominating mechanism to another.  

We consider the two-impurity Anderson model\cite{PhysRev.133.A1594} (TIAM) as 
one of the most basic models incorporating such competing interactions.
There, two magnetic impurities are immersed with finite distance
into a host with a noninteracting conduction band and hybridizations 
between conduction electrons and local interacting electrons.  

A single impurity is prone to the Kondo effect,\cite{hewson:KondoProblem93} 
where an effective antiferromagnetic exchange between 
the conduction electrons and the local spin of the magnetic impurity 
leads to a dynamic spin-screening. This archetypical 
many-body effect is associated with an characteristic
energy scale $T_\m{K}$ (Kondo scale) which is nonanalytic and 
exponentially small in terms of the effective exchange.

Additionally, the noninteracting band electrons induce an
effective magnetic exchange between the two impurities,\cite{PhysRev.96.99,*yosidaRKKY57,*kasuyaRKKY56}
known as the  Ruderman-Kittel-Kasuya-Yosida (RKKY)-interaction.
This RKKY-exchange can be ferromagnetic or antiferromagnetic, 
depending on the spatial separation 
of the two impurities.

The RKKY-exchange is expected to be dominant for a small 
hybridization between the impurity and the conduction band
and favors an inter-impurity coupling of the spins usually associated 
with long-ranged magnetic order.  The Kondo 
coupling, on the other hand, strongly  increases toward larger 
hybridization strengths and promotes individual Kondo effects
at each impurity resulting in two uncoupled and screened
impurities. 
This scenario proposed by Doniach\cite{doniachKLM77} for the competition 
between these two mechanisms is at the heart of the rich
variety observed in many physical 
systems.\cite{greweSteglichHF91,*colemanHeavyFermionMag07,*steglich:HFQCP06,*siQCHF10}

The TIAM as the simplest model for this competition has be studied intensively
\cite{jayaprakashTIK81,*PhysRevLett.58.843,*jones:twoImpurityKondo88,Sakai199081,*PhysRevB.35.4901,*PhysRevLett.72.916,*PhysRevB.59.85,*PhysRevB.65.140405,
  *Sakai1993323,*Nishimoto2006,*hattoriTIAM07,*mrossTIAM08,*leeTwoImp10,silvaTwoImpNRG96,zhuTIAM11} 
in the past and has been realized in a recent experiment.\cite{borkTwoImpKondo2011}
The thermodynamic properties of the model are governed by a quantum critical point separating the ordered
and paramagnetic (Kondo) ground states in  case of identical impurities and
particle-hole symmetry.\cite{PhysRevLett.58.843,*jones:twoImpurityKondo88}
However, the quantum critical point is unstable and turns into a crossover
once one of these symmetries is broken or  a direct hopping is included.\cite{Sakai199081,silvaTwoImpNRG96,zhuTIAM11} 
The dynamic properties under general conditions, that is varying impurity-impurity 
distance, hybridization and interaction strength, are not as clear.
Especially the situation with finite non-diagonal hybridization is 
not well studied since usually the RKKY exchange is simulated via
a direct exchange interaction.

In this work we present an approach to the a general TIAM which is formulated in the framework
of direct perturbation theory in the 
hybridization\cite{keiter:PerturbPRL70,*grewe:IVperturb81,*Grewe:lnca83,*keiterMorandiResolventPerturb84}
which utilizes  skeleton graphs for families of time ordered diagrams.
These techniques have  been laid out in the literature, e.g.\ in connection with the Kondo 
effect or the mixed valence problem,\cite{kuramoto:AnalyticsNCA85,*bickers:nca87a}
and have served to define approximation schemes of great usefulness and 
quality.\cite{greweCA108,*schmittSus09,kroha:NCA_CMTA_rev05}

There have already been  approaches to the TIAM with these techniques,\cite{schillerTwoImp93,*Schiller1996} 
but these were restricted to infinite Coulomb interaction, $U\to\infty$,  neglected vertex 
corrections, and ignored non-diagonal ionic propagators (see below for details).
The present scheme is an extension of the finite-$U$ enhanced noncrossing 
approximation (ENCA)\cite{pruschkeENCA89}, 
which improves on the well known noncrossing approximation (NCA)\cite{Grewe:siam83,*Kuramoto:ncaI83} 
by the inclusion of low-order vertex corrections.  It is also a direct extension
of the ENCA for the multi-orbital situation as described in 
Ref.~\onlinecite{greweENCACF09}.

Apart from an application to the rich and interesting physics encountered in the TIAM, 
our new two-impurity solver  also is of interest for extensions
of the dynamical mean-field theory to include  nonlocal two-site
correlations. Compared to other existing two-impurity solvers the 
present scheme has some advantages as well as 
shortcomings:  Like all the schemes based on perturbation theory in the 
hybridization, it has some problems to correctly describe the Fermi liquid properties 
at very low temperatures. This does not represent a major drawback as 
many scenarios and questions of current interest can be addressed at 
temperatures accessible within these schemes (see, for example, 
Ref.~\onlinecite{schmittNFLvHove10,*greteKink11,*schmittMagnet12}).
As will be demonstrated below and as it is known from the
single-impurity case,\cite{pruschkeENCA89,greweCA108,*schmittSus09,greweENCACF09}
these schemes correctly describe 
the non-perturbative Kondo-physics and are able to reproduce the
exponentially small low-energy Kondo scale. 
Additionally, there are no adjustable parameters 
and this theory is directly formulated 
for correlation functions at the real frequency axis
This avoids the inaccuracies connected with a numerical analytic
continuation of imaginary time data, as obtained, for example, from 
quantum Monte Carlo schemes.

\section{Model}
\label{sec:model}

We consider two impurities (labeled as $f$-electrons) placed at  
at lattice sites $\vek{R}_j$ with $j=\{1,2\}$.
That part of the Hamiltonian preserving all local occupation numbers
is
\begin{align}
  \hat H^f_{0}=&\sum_{j,\si} \eps^{f}_{j\si}\hat n_{j\si}^f
  +\sum_{j=\{1,2\}} U_{j} \hat n^f_{j\uparrow}\hat n^f_{j\downarrow}
  \notag \\   \label{eq:Hamf}
  &  +\frac{1}{2}\sum_{l\neq j, \si, \si'} U'_{12}\hat n^f_{j\si}\hat n ^f_{l \si'}.
\end{align}
The  operators $\f_{j\si}$ and $\fd_{j\si}$ are the usual  annihilation and  
creation operators for impurity $f$-electrons with spin $\sigma=\{+,-\}=\{\uparrow,\downarrow\}$
at the lattice $j$ and  $\hat n_{j\si}^f=\fd_{j\si}\f_{j\si}$. The density-density Coulomb 
interaction is incorporated via the matrix elements $U_j$ for on-site 
and $U'_{12}$ for inter-site interaction.  We also allow for more general couplings between 
the two impurities    
\begin{align}
  \label{eq:Wf}
  \hat W^f=&\sum_{l\neq j \si}t_{lj} \fd_{l\si} \f_{j\si}
  -\frac{1}{2}\sum_{l\neq j} J_{lj}\vek{\hat S}^f_l \vek{\hat S}^f_j+ \hat W'(X,Y),
\end{align}
where the first term is direct single-particle hopping with amplitude  $t_{lj}$
and the second term a Heisenberg exchange with exchange 
coupling $J_{lj}$ and $\vek{\hat S}_j^f$
the vector $f$-electron spin operator. The third term allows, e.g., 
correlated hopping (associated with matrix element $X$) and pair-hopping (associated with
matrix element $Y$).

These impurities  are immersed into a lattice with a noninteracting conduction 
band ($c$-electrons) with the dispersion relation  $\eps^c_{\vk}$,
\begin{align}
  \label{eq:Hamc}
  \hat H^c_{0}=&\sum_{\vk,\si}  \eps^c_{\vk}\hat  n_{\vk\si}^c
\end{align}
with the number operator for $c$-electrons  $\hat  n_{\vk\si}^c=\cd_{\vk\si}\c_{\vk\si}$
with spin $\sigma$ and crystal momentum $\vk$.
Both subsystems hybridize via a term 
\begin{align}
  \label{eq:V}
  \hat V=&\sum_{j,\vk,\si}\left(V_{\vk} e^{-i \vk\, \vek{R}_j}
    \cd_{\vk,\si} \f_{j \si}
    + h.c.\right).
\end{align}
The total Hamiltonian then reads
\begin{align}
  \label{eq:Hamtot}
  \hat H&=\hat H_0^f+\hat W^f+\hat H_0^c+\hat V
  .
\end{align}

Even though in principal there is no limitation in applying our calculational 
scheme to this model in full generality, in this work 
we will mostly consider identical impurities and focus on the half-filled
particle-hole symmetric case, where two electrons are placed in 
the two-impurity cluster.

\section{Novel two-impurity solver based on direct perturbation theory}
\label{sec:ii}

The general idea of direct
perturbation theory is to treat the two-impurity 
subsystem $\hat H_0^f+\hat H_0^c+\hat W^f$ exactly and to use $\hat V$ as
the perturbation. With $\hat W^f=\hat V=0$ the Fock-space of $f$-electrons is
diagonalized in terms of the local occupation numbers $\hat n^f_{j\si}$.
An eigenbasis of $\hat H_0^f$ is generated by
all product states with the factors taken from the two versions of the local
$f$-eigenbasis for the two sites $j=1$ and $2$ , respectively
\begin{align}
  \label{eq:basis}
  |0\rangle\phantom{{}_j}&\equiv   |n^f_{j\uparrow}=n^f_{j\downarrow}=0\rangle\\\notag
  |\!\!\uparrow\rangle_{j}&  \equiv |n^f_{j\uparrow}=1, n^f_{j\downarrow}=0\rangle=\fd_{j\uparrow}|{0}\rangle\\\notag
  |\!\!\downarrow\rangle_{j}&=\fd_{j\downarrow}|{0}\rangle\\\notag
  |2\rangle_j&=\fd_{j\uparrow}\fd_{j\downarrow}|{0}\rangle
\end{align}
The inclusion of the term $\hat W^f$  makes it necessary to transform to 
a different many-electron eigenbasis of $f$-states  in order to diagonalize the 
two-impurity subsystem. 
The following discussion of the novel
two-impurity solver is formulated in general terms. It uses the afore mentioned
multi-electron
eigenstates of the diagonalized $f$-system, 
\begin{align}
  \label{eq:24}
  \hat H_0^f|M\rangle&= E_M|M\rangle,
\end{align}
labeled by quantum numbers $M$.
The corresponding Hubbard-transfer operators 
\begin{align}
  \label{eq:XOp}
  \hat X_{M M'}&=|M\rangle\langle M'|
\end{align}
allow for the decomposition of one-particle creation- and annihilation
operators (with coefficients specified later),
\begin{align}
  \label{eq:25}
  \f_{j\si}&=\sum_{\substack{M,M'}} \alpha^{j\si}_{M M'}  \hat X_{M M'},\\
  \notag
  \fd_{j\si}&=\sum_{\substack{M,M'}} \left(\alpha^{j\si}_{M M'}\right)^*  \hat X_{M' M}
  .
\end{align}

In direct
perturbation theory the partition function and the fermionic 
two-point Green functions are expressed as contour integrals in the complex
energy plane:\cite{keiter:PerturbPRL70,*grewe:IVperturb81,*Grewe:lnca83,*keiterMorandiResolventPerturb84}
\begin{align}
  \label{eq:Zloc}
  \partition&=\oint_{\mathcal{C}}\frac{dz}{2\pi i} e^{-\beta z}
  \mathrm{Tr}\left[\mathcal{\hat R}(z)\right]
    \\ \notag
    &=\partition_c   \sum_{M}\int\limits_{-\infty}^{\infty} d\omega e^{-\beta \omega} \left[-\frac{1}{\pi} \Im P_{M;M}(\omega+i\delta)\right],
  \end{align}
  \begin{align}
    \label{eq:26}
    &G_{\scriptscriptstyle{\hat A,\hat B}}(i\om_{n})=\frac{1}{\partition}
    \oint\limits_{\substack{\mathcal{C}}}\frac{dz}{2\pi i} e^{-\beta z}
    \mathrm{Tr}\left[\mathcal{\hat R}(z)\; \hat A\; \mathcal{\hat R}(z+i\om_n)\;\hat B\right]
    \notag     \\
    &\phantom{M}= \frac{1}{\partition_f}
    \sum_{\substack{M',M'',\\M''',M}} \alpha_{M',M''}^{A} \alpha_{M''',M}^{B}\oint_{\mathcal{C}}\frac{dz}{2\pi i} e^{-\beta z} 
    \times
    \\\notag
    &\phantom{M}\times
    P_{\scriptscriptstyle{M;M'}}(z)\Lambda_{\scriptscriptstyle{M';M''}}(z,z+i\om_n)P_{\scriptscriptstyle{M'';M'''}}(z+i\om_n).
\end{align}
with the resolvent operator $\mathcal{\hat R}(z)=\left[z-\hat H\right]^{-1}$ and 
$\mathcal{Z}_c=\mathrm{Tr}_{c}\left(e^{-\beta \hat H_c}\right)$ is the partition
function of the noninteracting subsystem of band electrons. 
The $P_{M,M'}(z)$ are so-called ionic propagators which capture the correlated 
dynamics of a transition from an ionic $f$-state with quantum number $M'$ to a state with $M$.
These are exact equations if  the exact vertex functions $\Lambda_{\scriptscriptstyle{M';M''}}(z,z+i\om_n)$
are used.
In practice, however,  the exact vertex functions are unknown and approximations for 
$\Lambda_{\scriptscriptstyle{M';M''}}$ are employed.

For simplicity, we will confine our considerations to one-particle $f$-Green functions
where $\hat A=\f_{j\sigma}$ and $\hat B=\fd_{l\sigma'}$. 
For the two-impurity model the one-particle Green function is conveniently written 
as a $2\times 2$ matrix for each spin component, 
\begin{equation}
  \ull{G}_{\si}=\left(
    \begin{array}{cc}
      G_{\f_{1\si},\fd_{1\si}} &     G_{\f_{1\si},\fd_{2\si}} \\
      G_{\f_{2\si},\fd_{1\si}} &     G_{\f_{2\si},\fd_{2\si}} \\
    \end{array}\right). 
  \label{eq:27}
\end{equation}
A central quantity of physical interest is the one-particle spectral function
which is given by  the diagonal part of the Green function,      
\begin{align}
  \label{eq:rhof}
  \rho^f_j(\omega)=-\frac 1\pi\mathrm{Im} G_{\f_{j\si},\fd_{j\si}}(\omega+i\delta)
  ,
\end{align}
where the limit $0<\delta \to 0$ is implied.

In contrast to what is commonly needed for (effective) single-impurity
models,\cite{Grewe:siam83,*Kuramoto:ncaI83,greweCA108} we introduced  non-diagonal ionic 
propagators,
\begin{equation}
  P_{M;M'}(z)=\sum_{\substack{\mathrm{conduction}\\\;\; \mathrm{states}\; \gamma}}\frac{e^{-\beta E_{\gamma}}}{\mathcal{Z}_c}\langle\gamma,M|\mathcal{\hat R}(z+E_{\gamma})|M',\gamma\rangle
  \label{eq:28}
\end{equation}
which together with the diagonal components and the  vertex functions $\Lambda_{M;M'}(z,z')$ are
the central quantities of interest which need to be computed.
 
The perturbational processes contributing to the ionic propagators and vertex
functions take place along the imaginary time axis $0\leq \tau < \beta
\equiv \frac{1}{T}$. 
These processes can be partially summed-up to
families of time-rotated diagrams on cylinders where the  endpoints $\tau=0$
and $\tau=\beta$ are identified. The Green function of Eq.\ (\ref{eq:26})  
is represented by one particular diagram where the external line 
(accounting for the operator $\hat B$) is attached to a bare vertex 
occuring at the latest time; the dressed vertex 
$\Lambda_{M,M'}(z,z')$, where the external line  associated with the operator 
$\hat A$ is attached, occurs at an earlier time. The initial and final ionic state $M$ is 
displayed twice  in graphical representations for clarity, but is included
only once in the corresponding formulas.

As usual, propagators are favorably built up from irreducible
self-energies $\Sigma_{M;M'}(z)$, which together with the vertex functions form a
closed set of skeleton equations. Diagrammatic contributions to the
self-energies are identified as pieces of diagrams which cannot be
separated into disjoint parts by cutting only the time axis of each
site in the cluster.

It is important to realize here, that the
time-ordered processes, generated by expanding the resolvents in
Eq.\ (\ref{eq:28}) in terms of the hybridization $\hat V$, 
generally extend over all sites of the cluster. 
Local pieces on one particular site $j$ can occur in proper time-order, either
after an electron has is transferred from a different site $l$
 or after an interaction process, both induced by  $\hat W^f$. 
If $\hat W^f$ would have 
been regarded as a perturbation, too, these events would appear in the diagrams
explicitly. However, we will take $\hat W^f$ into account exactly via a
non-diagonal form of the ionic propagators $P^{(0)}_{M; M'}(z)$, which
corresponds to the unperturbed Hamiltonian $\hat H^f_0+\hat H^c_0+\hat W^f$.
  
As a consequence of the all-embracing unique time-order in the cluster only
one energy variable is present after the Laplace-transform. Another
consequence of this fact concerns the dynamical interdependence of local
parts of a process in a cluster. 
As already mentioned, all events occur in a definite time-order. Even if two
processes happen to be located at separate impurities, 
e.g.\ the excitation of two band electrons and re-absorption at the respective same site,
different time-orderings of the individual two excitations and absorptions are
counted separately in this approach. For example, interchanging the emission times of the two
processes generates or removes one crossing of two band-electron lines.      
As the approximation is organized in numbers of crossing band electron lines
higher-order vertex corrections with a multitude of crossing lines are not included.
This introduces artificial inter-site correlations. 

Therefore, all such  approximations within a two-impurity model 
will in principle not faithfully reduce to two independent sites in the case 
of uncoupled impurities. They have to be thoroughly checked 
in the limiting case of two independent sites, where an ordinary Kondo effect
is expected at each site. As it turns out, the two-impurity solvers to be introduced below
are capable of describing this limiting case reasonably well.
On the other hand, these  approximation schemes
will perform even better when real physical inter-site correlations 
make it necessary to respect overall time order.

The connection between the irreducible self-energies and the 
ionic propagators is provided by a Dyson-equation  
\begin{align}
  \label{eq:29}
  P_{M;M'}(z)=&P^{(0)}_{M;M'}(z)\\
  \nonumber
  &+\sum_{M'',M'''}P^{(0)}_{M;M''}\Sigma_{M'';M'''}(z)
    P_{M''';M'}(z)
    .
\end{align}
As already stated above, the action of $\hat W^f$ is included in $P^{(0)}_{M;M'}$, i.e.\
it is a solution to the equation 
\begin{align}
  P^{(0)}_{M;M'}(z)&=\frac{1}{z-E_M}\delta_{M,M'}\\\notag
  &\phantom{=}+\frac{1}{z-E_M}\sum_{M''}\langle\hat W^f\rangle_{M,M''}^{\phantom{{(0)}}}
  P^{(0)}_{M'';M'}(z).
\end{align}
The irreducible self-energy $\Sigma_{M;M'}$ incorporates the 
additional modifications brought about by the hybridization $\hat V$.
The limit of an isolated cluster not hybridizing with the conduction band
is thereby exactly reproduced 
for vanishing hybridization with the band states, since
\begin{equation}
  \label{eq:30}
  \qquad V_{\vk}\to  0:\quad \Sigma_{M;M'}(z) \to 0 
  .
\end{equation}

For the present case of a cluster with two sites $j=1,2$ and 
$s$-shells only, the quantum numbers are  specified as
$M=\left(m_1,m_2\right)$ with $m_j=\{0, \uparrow, \downarrow, 2\}$.
The  annihilation operators of Eq.~\eqref{eq:25} are then given as
\begin{align}
  \label{eq:31}
  \f_{1\si}=&\big[|0\rangle_1{}_1\!\langle\si|+\si|-\si\rangle_1{}_1\!\langle 
  2| \big]\otimes\hat 1_2,
  \\\notag
  \f_{2\si}=&\hat 1_1\otimes \big[|0\rangle_2{}_2\langle\si|+\si|-\si\rangle_2{}_2\!\langle 2| \big],
  \\\notag
  \hat 1_j=&|0\rangle_j{}_j\langle 0|+(-1)^j
  \sum_{\substack{\sigma}}|\si\rangle_j{}_j\!\langle\si|+|2\rangle_j{}_j\!\langle
  2|.
  \end{align}
The differing sign in the definition of the local identity operators 
for impurity 1 and 2 stems from the definition of the basis states,
in particular from the prescribed order in which the creation operators
act to produce the other basis state from the two-site vacuum.

The overall sign of a diagram turns out to be more subtle for the 
two-impurity case than for a single-impurity. In the latter 
the overall sign is determined by simply including a factor of $(-1)$
for each crossing of two band electron lines. For the two-impurity case,
we have found no better way than to count all exchange processes of 
Fermi-operators in the course of reducing expectation values of operator 
products to normal order. This is done
algorithmically on a computer along with the enumeration and
classification of all different diagrams contribution to the quantity of interest.

\begin{figure}[t]
  \centering 
  \includegraphics[width=0.4\linewidth]{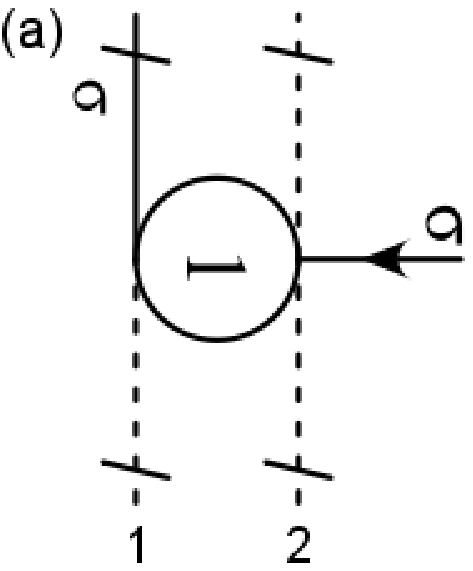}
  \includegraphics[width=0.4\linewidth]{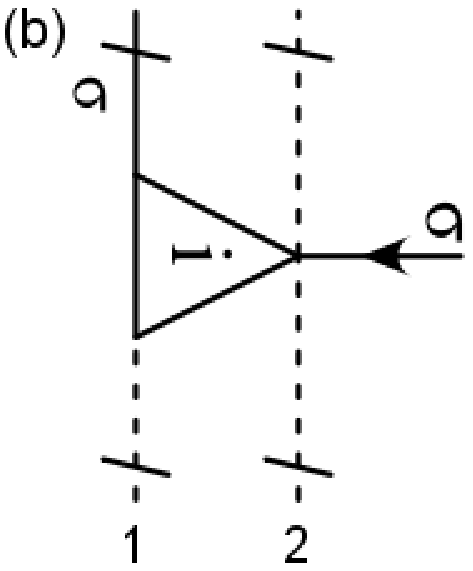}
  \caption{(a) Bare vertex depicting an absorption 
    of a band electron  with spin $\sigma$ at site 1,
    changing the two-impurity quantum number from $M=(0,0)$ 
    to the  $M'=(\sigma,0)$. 
    (b) Dressed vertex for the absorption 
    of a band electron  with spin $\sigma$ at site $i$.
    The left (right) vertical line denotes site 1 (site 2)
    and the incoming arrow a band electron.
    A dashed vertical line represents an empty and a full vertical
    line a singly occupied shell, where 
    the spin of the electron is indicated next to the line. 
    The number in the vertex symbol denotes the site, to which the band electron line 
    is attached. For the bare vertex this can occur only on site 1,
    while for the dressed vertex this is not specified.
    In an analytic expression, the bare vertex amounts to a factor of $1$, while
    the dressed vertex stands for the function $\Lambda_{00;0\si}^i(z,z+\eps)$,
    when the band electron has an energy $\epsilon$ and the local two-site cluster
    has the complex energy variable $z$.
  }
  \label{fig:4}
\end{figure}

The analytic contributions are visualized in
a diagrammatic language similar to the one developed long ago for the
single-impurity problem,\cite{keiter:PerturbJAP71} however, with a few characteristic
differences. 
Even though there exists only one energy  variable $z$ for both sites,
we distinguish two vertical lines corresponding to local
processes on sites 1 (left) and 2 (right) for clarity.
These become decorated with vertices ordered along the
vertical direction
due to hybridization events. Bare vertices are depicted 
as circles with the corresponding number of the site where the 
external line is attached at,
and dressed vertices are depicted as triangles; they  embrace both
lines as shown in Fig.\ \ref{fig:4}(a) and (b).

\begin{figure*}[ht]
  \centering
  \includegraphics[width=0.8\linewidth]{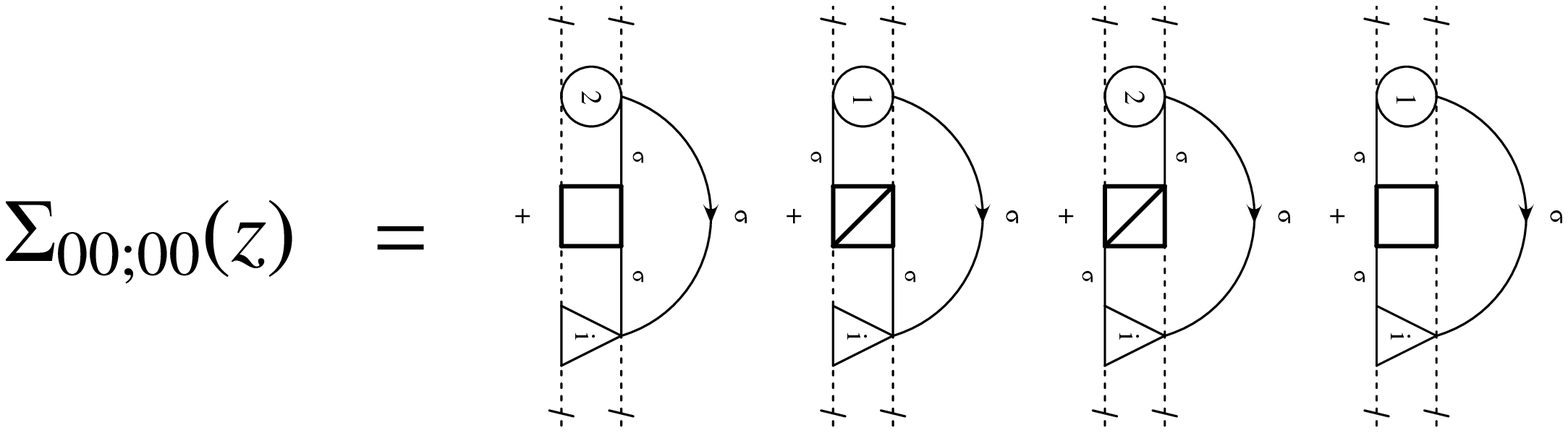}
  \caption{Contributions to the self-energy $\Sigma_{00;00}(z)$. 
    The squares represent ionic propagators,
      $P_{m_{f1}m_{f2}; m_{i1}m_{i2}}(z)$,
      where the earlier [later] state characterized by the quantum numbers $M=(m_{i1}, m_{i2})$ 
      [$M=(m_{f1}, m_{f2})$] is indicated by the attached lower [upper] two lines.
      For clarity, a diagonal line inside the box is added in case of a non-diagonal propagator. 
      Summations over all internal quantum numbers  is understood in all  graphs presented in this work. 
    }
  \label{fig:5}
\end{figure*}

The ionic propagators between vertices are nonlocal  objects 
and non-diagonal in quantum numbers $M$, and as such they  
describe the combined time evolution of both ionic
states. Band-electron lines are always considered to be external to the 
whole cluster and can be tied to either of both sites  indicated by the number
inside the vertex symbol. The short horizontal lines 
denote amputation as usual, where the initial and final states are not included in
the analytical representations. The translation of a diagram into an
analytical expression is handled in complete analogy to the
diagrammatic rules stated in the 
literature,\cite{keiter:PerturbPRL70,*grewe:IVperturb81,*Grewe:lnca83,*keiterMorandiResolventPerturb84,kuramoto:AnalyticsNCA85,*bickers:nca87a}
except for the sign rule mentioned above. 

Hybridization events occur as diagonal ($j=l$) and 
non-diagonal ($j\neq l$) processes characterized by 
the (complex) functions
\begin{equation}
  \label{eq:Gamma}
  \Gamma_{j,l}(z)=\frac{1}{N}\sum_{\vk}\frac{|V_{\vk}|^2 e^{i \vk\;\left(\vek{R}_l-\vek{R}_j\right)}}{z-\eps_{\vk}}
  .
\end{equation}
Their imaginary part at the real frequency axis ($z=\omega\pm i\delta$)
\begin{align}
  \label{eq:32}
  \Delta_{j,l}(\omega)&=\frac 1{2i}\big[\Gamma_{j,l}(\om-i\delta)-\Gamma_{j,l}(\om+i\delta)\big]\\\notag
  &=\frac{1}{N}\sum_{\vk}|V_{\vk}|^2 e^{i \vk\;\left(\vek{R}_l-\vek{R}_j\right)}\pi \delta(\om-\eps_{\vk})
\end{align}
is called hybridization function and is the physically most relevant part. 
The diagonal component at zero frequency corresponds to the well-known 
Anderson width of a resonant impurity level, $\Delta\equiv\Delta_{jj}(\om=0)$. As with the other quantities,
the diagonal and non-diagonal components are conveniently 
treated as  matrices.

We  refrain from presenting the complete system of self-energy
and vertex-equations, which even after using all symmetries of the
two-site Anderson-model, is far too large to present (It can be
found, however, in Ref.~\onlinecite{Jabben2010}.) 
Instead we exemplify the structure by selected 
examples for  self-energy and  vertex equations.

\begin{figure*}
  \includegraphics[width=0.9\linewidth]{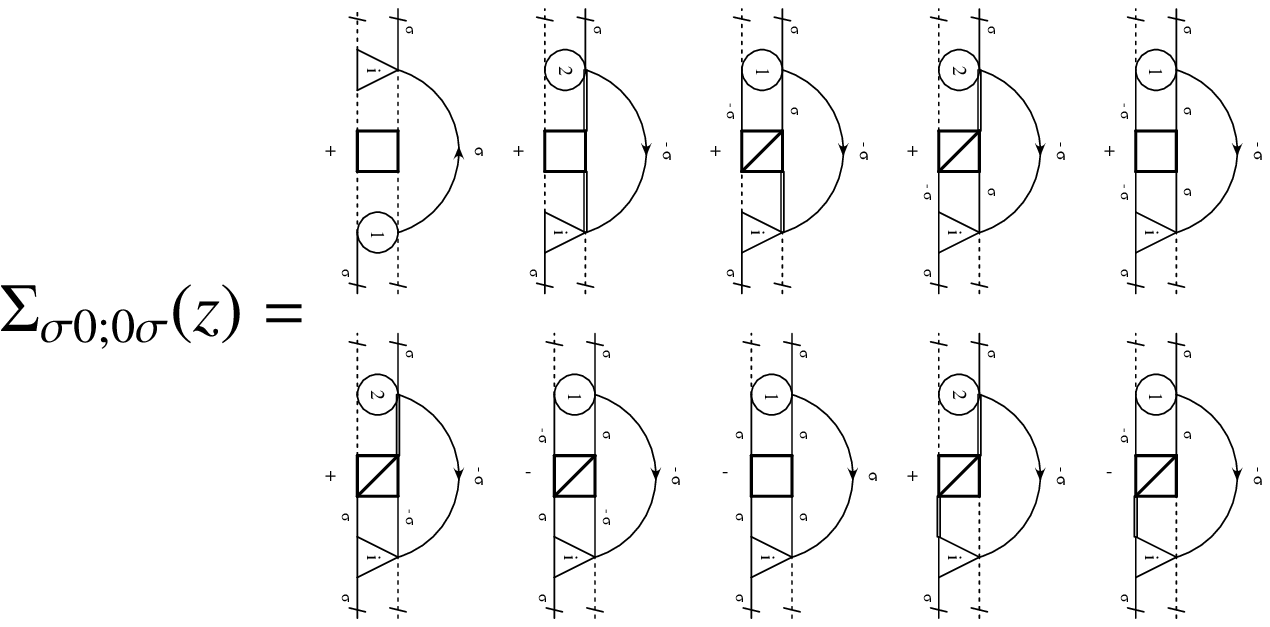} 
  \caption{Contributions to the self-energy $\Sigma_{\si 0;0\si}(z)$.
    The double-line visible in some contributions indicates a doubly occupied 
    ionic shell on one site.
  }
  \label{fig:6}
\end{figure*}

The cluster vacuum with no electrons on each of the two ionic shells
experiences excitations according to the self-energy diagrams shown in
Fig.~\ref{fig:5}. 
The corresponding analytic contribution is
\begin{align}
  \label{eq:33}
  \Sigma_{00;00}(z)&=\sum\limits_{i,\si}
  \int\limits_{-\infty}^{\infty} \frac{d\eps}{\pi}\Big[
  \\\notag
  & 
  \phantom{-}\Delta_{i,2}(\eps) f(\eps)\; \Lambda_{00;0\si}^i(z,z+\eps)
  \;      P_{0\si;0\si}(z+\eps)
    \\\notag
    &+ 
      \Delta_{i,1}(\eps) f(\eps) \;\Lambda_{00;0\si}^{i}(z,z+\eps)
      \;     P_{0\si;\si0}(z+\eps)
      \\\notag
    &+
      \Delta_{i,2}(\eps) f(\eps)\; \Lambda_{00;0\si}^{i}(z,z+\eps)
      \;P_{\si0;0\si}(z+\eps)
      \\\notag
    &+
      \Delta_{i,1}(\eps) f(\eps)\; \Lambda_{00;0\si}^{i}(z,z+\eps)
      \;P_{\si0;\si0}(z+\eps) \Big]
      .
\end{align}
Among the graphical elements in Fig.~\ref{fig:5} is an  
insertion with a diagonal line inside the square. This  represents a non-diagonal
ionic propagator where an electron is transferred from one site of the
cluster to the other. 
Figure \ref{fig:6} presents the diagrammatic contributions to 
a non-diagonal self-energy. 

Up to now, all the equations for propagators and self-energies are exact. However, 
the  dressed vertex functions, as indicated by the triangles in the graphs, have not been
specified yet. 
Clearly,  the exact vertex functions
are in general unknown and approximations have to be introduced. 

The simplest approximation amounts to ignoring all vertex corrections
and just replacing every full vertex by a bare one, that is $\Lambda_{M;M'}^j\to 1$.
The result is the natural generalization
of the NCA for finite $U$ (usually termed SNCA) to the two-impurity case
and will be referred to as two-impurity SNCA.

\begin{figure*}[ht]
  \includegraphics[width=\linewidth]{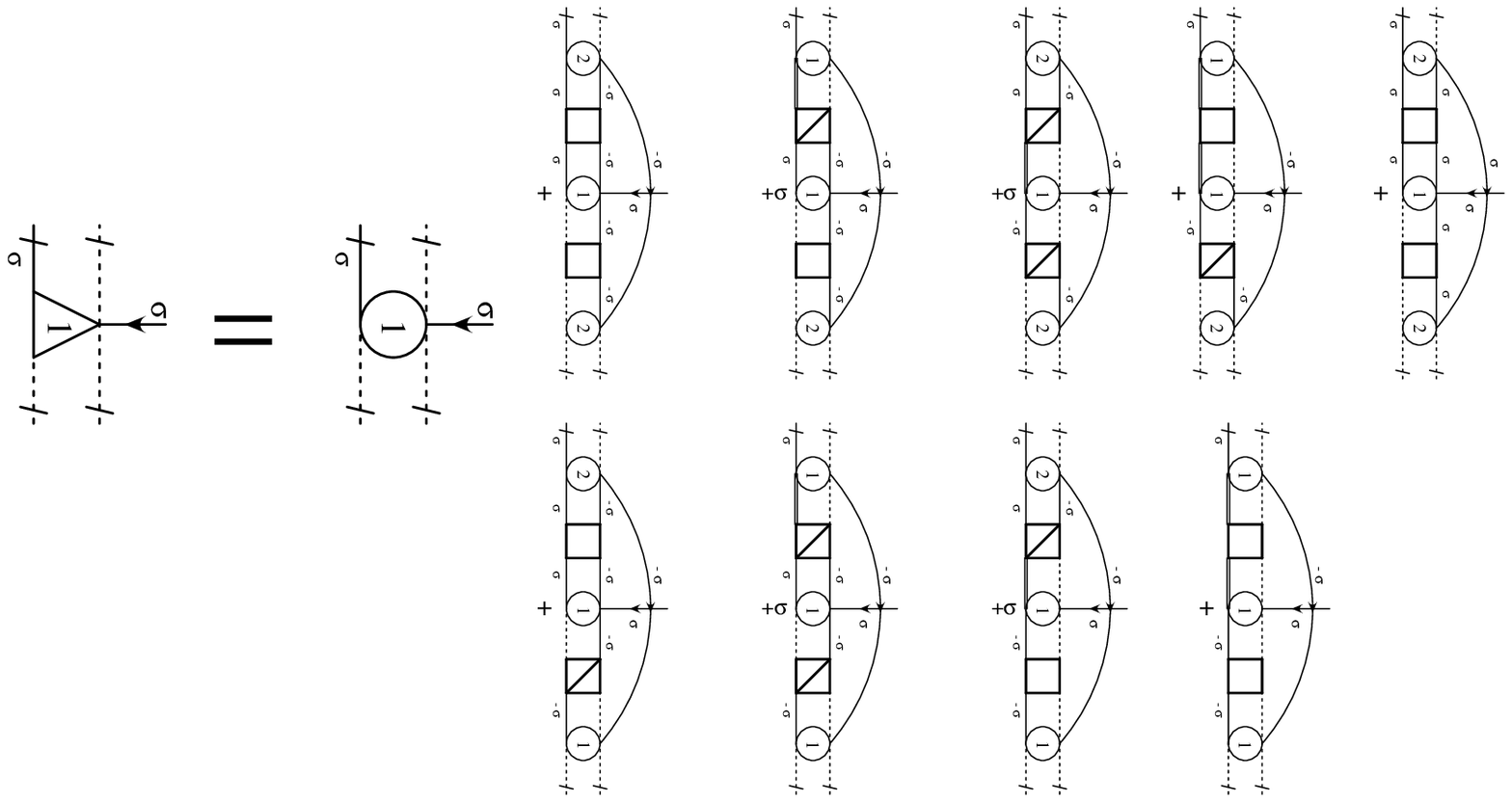}
  \caption{Contributions to the vertex $\Lambda_{00;\sigma0}^1$
    within the two-impurity ENCA approximation which includes the first oder of 
    vertex corrections.}
  \label{fig:7}
\end{figure*}

In order to incorporate
the correct order of magnitude for the Schrieffer-Wolff exchange coupling  for finite $U$,
the first order vertex corrections with one crossing of band electron lines have
to be included.\cite{pruschkeENCA89} 
The extension of this ENCA for the single-impurity  to 
the present two-impurity set-up will be termed two-impurity ENCA.   
As an example we present the vertex function 
$\Lambda_{00;\sigma0}^i(z,z')$ within the two-impurity ENCA in Fig.\ \ref{fig:7}.

A closed system of coupled equations for ionic propagators, self-energies and vertex functions
includes many such integral equations. Great care is needed to identify all
different processes together with the proper assignment of quantum numbers, energies,
matrix elements and signs.
This coupled system is solved self-consistently for the unknown 
self-energies and ionic propagators. There are between
6 and 158 different ionic propagators, depending on symmetries,
and about 10-100 different diagrammatic contributions to each self-energy.
Therefore, the number of necessary convolutions in each iteration step
is easily in the thousands. As in the usual procedure, 
an additional set of coupled integral equations has 
to be solved for the so-called defect propagators.\cite{Kuramoto:ncaIII84,greweCA108,*schmittSus09}

While the computational effort for the two-impurity schemes presented here is 
increased considerably compared to their single-impurity counterparts, the 
numerical difficulties are of the 
same nature:
The ionic propagators exhibit very narrow features near the ionic thresholds which
can be handled by the  usage of optimally adapted energy meshes. 
The strongest challenge is posed by  the numerical solution of the system 
of equations for the defect propagators. The lack of an absolute scale, i.e.\
the lack of the knowledge of the partition function, leads to 
suboptimal convergence properties.\cite{schmittPhD08,Jabben2010}
However, proper and rather accurate 
solutions can indeed be obtained, and results will be presented in the following 
sections.
The interested reader can consult Refs.~\onlinecite{Jabben2010}
and \onlinecite{schmittPhD08} for more details.

When extending these approximations to systems with more local orbitals or
impurities 
one is confronted with the fundamental problem of an
exponentially  increasing number  of local many-body states.
Additionally, 
the inclusion of vertex corrections as in the two-impurity ENCA further
increases the 
contributions to each ionic self-energy.
However, in the present approach symmetries can easily be implemented by 
identifying equivalent propagators; thus, the computational cost is 
dramatically reduced. In this respect it is also very effective and easily 
implemented to restrict the allowed local occupation and to exclude many-body 
states with very high excitation energies, as it was done for the two-orbital 
Anderson model (Ref.~\onlinecite{greweENCACF09}).

\section{Results for uncoupled impurities}

In this section we first concentrate on the basic question
of how the novel two-impurity solver   performs in 
some simple model situations. 
We compare the  two-impurity SNCA and ENCA with
the corresponding SNCA- and ENCA-solutions of the 
single-impurity Anderson problem in order to check on the 
presence and relative weight of
artificial correlations introduced by the two-impurity treatment.

We assume spin-degeneracy in the following  and measure energies   with 
reference to the chemical potential $\mu=\omega=0$. We always take
the noninteracting conduction electrons to have the
tight-binding dispersion of a three-dimensional simple-cubic 
lattice 
\begin{align}
  \label{eq:dispersion}
  \epsilon_{\vek k}^c=-2t\big[\cos(k_x)+\cos(k_y)+\cos(k_z)\big]
\end{align}
(with the lattice spacing set to one, $a=1$).
As the unit of energy we take either the half-bandwidth of the noninteracting 
conduction band, $D=6t$, or the Anderson width $\Delta$, depending on what
seems more appropriate. 
We also restrict ourselves to identical impurities, that is 
$\epsilon^f_1=\epsilon^f_2=\epsilon^f$, $U_1=U_2=U$,
and $\rho^f\equiv\rho^f_j$
and mostly focus on the two-electron sector.

We define even ($+$) and  odd ($-$) parity one-particle 
states\cite{jayaprakashTIK81}
\begin{align}
  \label{eq:2}
  \f_{\pm\si}=\frac{1}{\sqrt{2}} ( \f_{1\si}\pm \f_{2\si})
\end{align}
and also introduce even and odd hybridization functions,
\begin{align}
  \label{eq:2a}
  \Delta_{+}(\omega)&=\frac{2}{N}\sum_{\vk}|V_{\vk}|^2\cos^2\big(\frac{1}{2}\vk\; \vek{d}\big)\pi\delta(\omega-\eps^c_{\vk})\\
  \Delta_{-}(\omega)&=\frac{2}{N}\sum_{\vk}|V_{\vk}|^2\sin^2\big(\frac{1}{2}\vk\; \vek{d}\big)\pi\delta(\omega-\eps^c_{\vk})
  .
\end{align}
Here, $\vek{d}=\vek{R}_1-\vek{R}_2$ is the distance vector  
between the two impurities.
For very large distances, $|\vek d|\to \infty$,
$\Delta_{\pm}(\omega)$ both approach the single-impurity
hybridization function,
\begin{align}
  \label{eq:siamHyb}
  \Delta^{\mathrm{SIAM}}(\omega)&=\frac{1}{N}\sum_{\vk}|V_{\vk}|^2\pi\delta(\om-\eps^c_{\vk})
  .
\end{align}
For vanishing distance, $|\vek d| \rightarrow 0$,
$\Delta_+(\omega)=2\Delta^{\mathrm{SIAM}}(\omega)$ and $\Delta_-(\omega)=0$.

\begin{figure}[ht]
  \centering
  \includegraphics[width=0.9\linewidth]{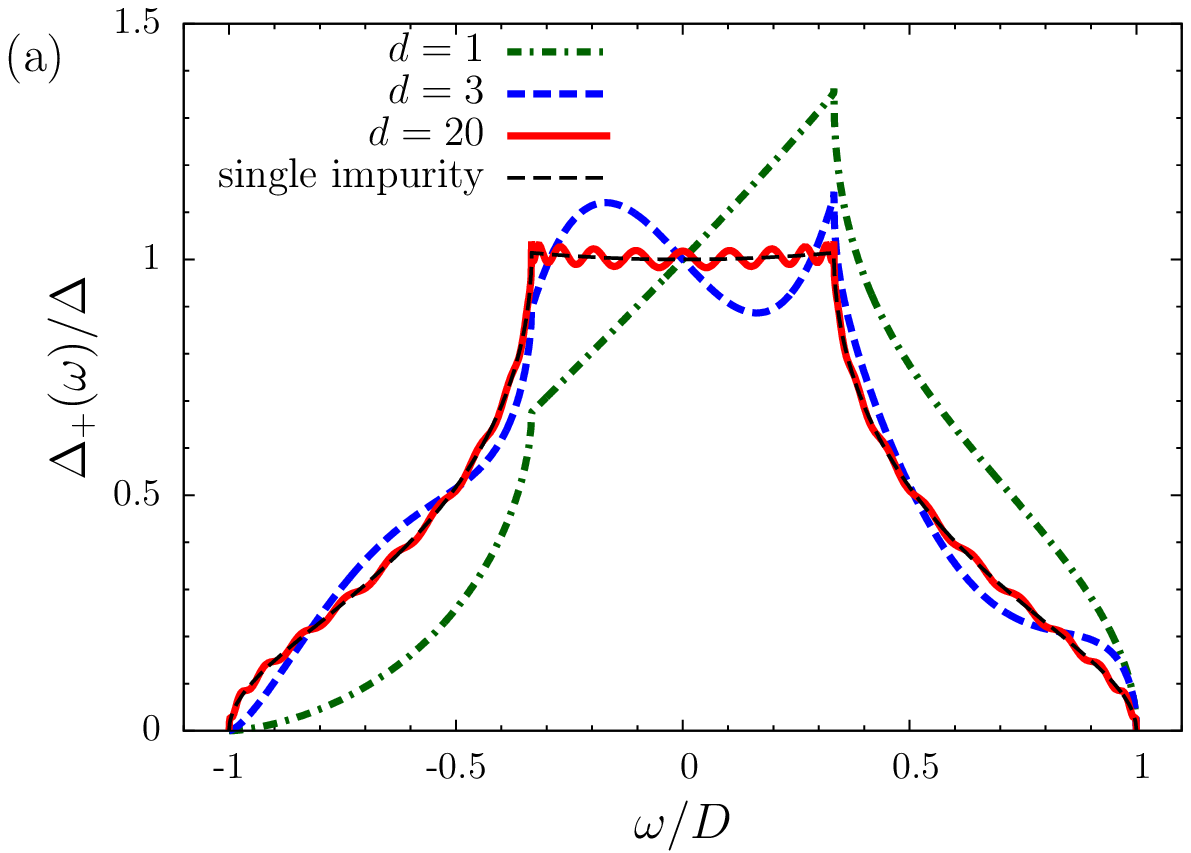}
  \includegraphics[width=0.9\linewidth]{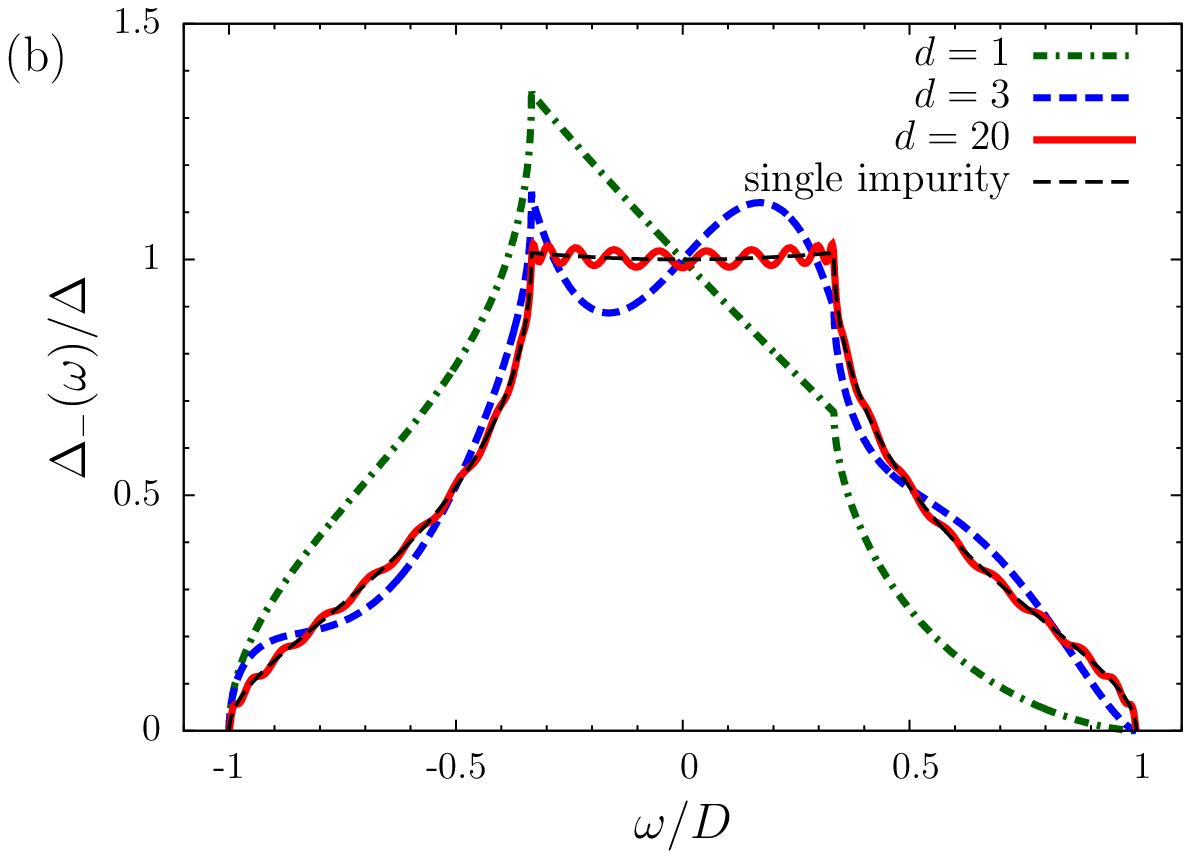}
  \caption{(Color online) Hybridization functions normalized to the Anderson width $\Delta$
    for the even (a) and odd (b) channel for various impurity-impurity
    distances $d=|\vek d|$ and momentum independent hybridization matrix element, 
    $|V_{\vek{k}}|^2\equiv|V|^2$.
    For comparison the  single impurity hybridization function is also shown, 
    which is nothing but the rescaled
    density of states of the  noninteracting band electrons.}
\label{fig:res1}
\end{figure}

These even and odd hybridization functions are shown in Fig.~\ref{fig:res1} 
for the three-dimensional tight-binding bandstructure of a simple-cubic lattice 
and the distance vector along a principal direction, e.g.\ $\vek d=d\vek e_x$.
We also assumed a constant hybridization matrix element, $V_{\vk}=V$.

The first limiting case we like to investigate is obtained if 
all couplings between the impurities 
of the cluster are suppressed, i.e.\ $\Delta_{12}=0$, and likewise all matrix elements 
in the Hamiltonian which directly couple the two impurities, 
i.e.\  $U_{12}'=0$ and $\hat W^f=0$. 

Although the resonant level limit $U=0$
is trivially solved exactly, the reconstruction of this solution via
direct perturbation theory is highly nontrivial.\cite{Grewe:siam83,schmittPhD08} 
In fact, this case turns out to be particularly unfavorable for 
direct perturbation theory.

The spectral densities of even and odd states in the case of $U=0$
can be easily  derived from the exact one-particle Green functions 
\begin{align}
  \label{eq:3}
  \rho^f_{\pm}(\om)&=-\frac{1}{\pi} \mathrm {Im} G_{\pm}(\om+i\delta)\\
  G_{\pm}(z)&=\frac1{z-\eps^f-\Gamma_{\pm}(z)-\frac{g(z)^2}{z-\epsilon^f-\Gamma_{\mp}(z)}}
\end{align}
where generally an additional coupling function
\begin{align}
  g(z)=\frac{1}{N}\sum_{\vk}\frac{|V_{\vk}|^2\sin(\vk\; \vek{d})}{z-\eps^c_{\vk}}
\end{align}
occurs.
However, for the present case of a momentum independent hybridization matrix element, 
$V_{\vek k}=V$,
and inversion symmetric lattices where $\eps^c_{-\vk}=\eps^c_{\vk}$,
this function vanishes, $g(z)=0$.

\begin{figure}[ht]
  \centering
  \includegraphics[width=0.9\linewidth]{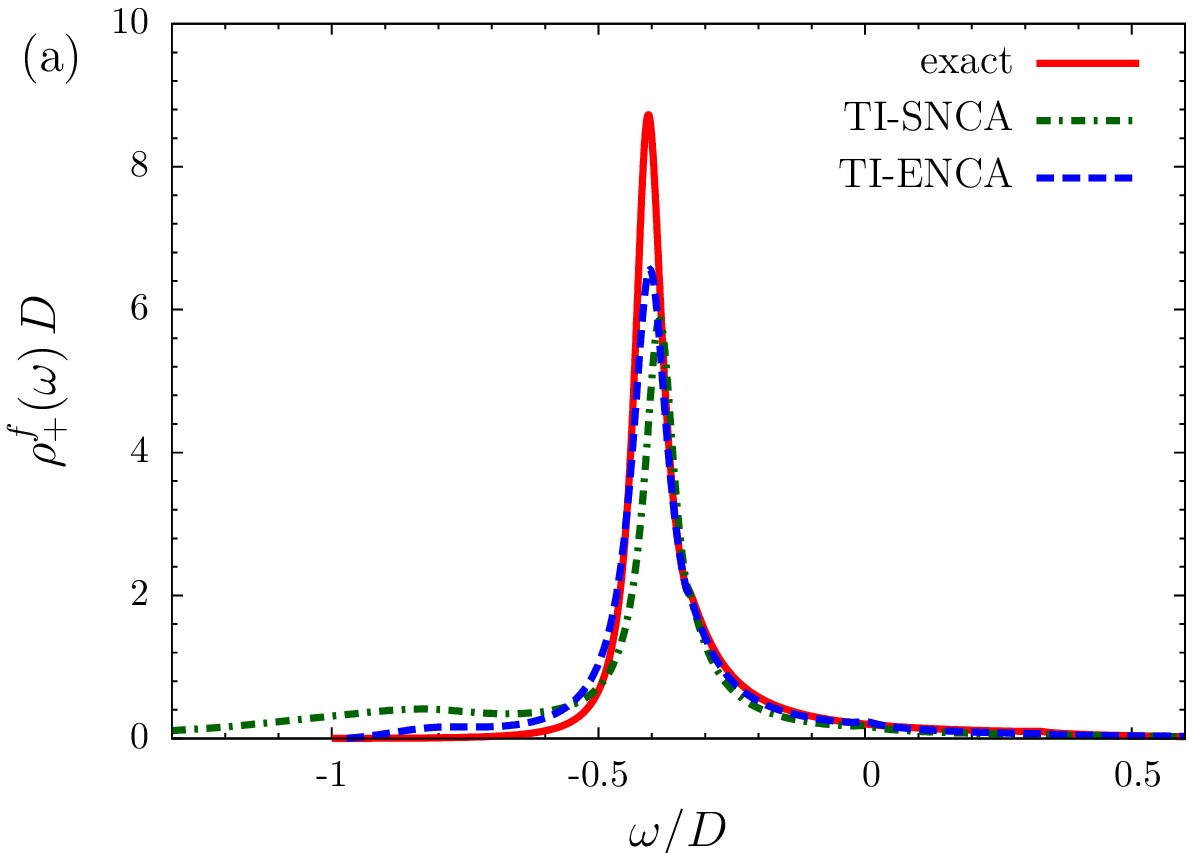}
  \includegraphics[width=0.9\linewidth]{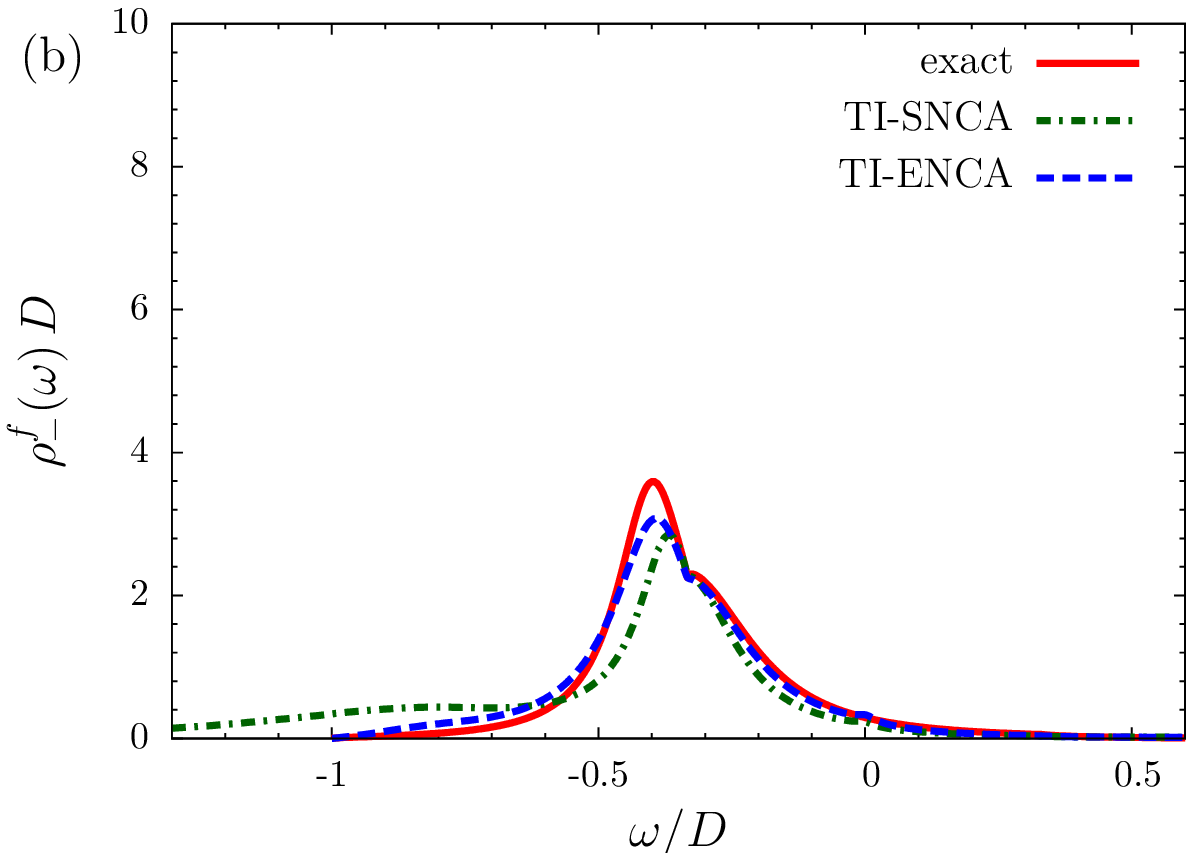}
  \caption{(Color online) Exact spectral functions of the resonant level with 
    $\epsilon^f=-0.4D$ [see  Eq.~\eqref{eq:3}]
    in the even (a) and odd (b) channel for distance $d=1$
    compared to the  two-impurity (TI-) SNCA and ENCA results.
  }
  \label{fig:res2}
\end{figure}
The exact curves for $U=0$, $\epsilon^f=-0.4D$ and distance $d=|\vek d|=1$ are 
compared with the results
of the two-impurity SNCA and two-impurity ENCA calculations 
in Fig.~\ref{fig:res2}.
Although the overall features of all curves show rough 
agreement, the two approximations perform differently in detail.
Both capture the van-Hove singularity on the 
right flank of the resonance at $\omega/D=-\frac 13$, 
but the two-impurity ENCA comes closer to the exact solution
than two-impurity SNCA. The position and height of the maximum is also   captured better
by the two-impurity ENCA. Additionally, the two-impurity SNCA produces 
some unphysical weight
in the left flank of the peak at $\omega/D\approx 0.75$.

As mentioned in the previous section
such deficiencies had to be expected.
They turn out, however, to be less pronounced as one might have
feared, so that even in this most critical case of uncorrelated impurities 
a reasonable result is produced by the two-impurity ENCA. As in the single impurity case, 
the performance should even increase when correlations become important for $U>0$. 

We  now turn to the Kondo regime with $U\gtrsim-2\eps^f > \Delta$.
The hybridization $\hat V$ tends to produce separate Kondo effects at each impurity
controlled by the diagonal elements of the  hybridization functions $\Delta_{jj}(\omega)$,
but additionally, the non-diagonal element of the hybridization, $\Delta_{lj}(\omega)$ (with $l\neq j$),
will produce a RKKY-interaction 
between the local magnetic moments on the impurities. In order to benchmark
the approximations we neglected the latter for the moment and set $\Delta_{lj}(\omega)=0$ for $l\neq j$.

\begin{figure}[ht]
  \centering
  \label{fig:3a}\includegraphics[width=0.9\linewidth]{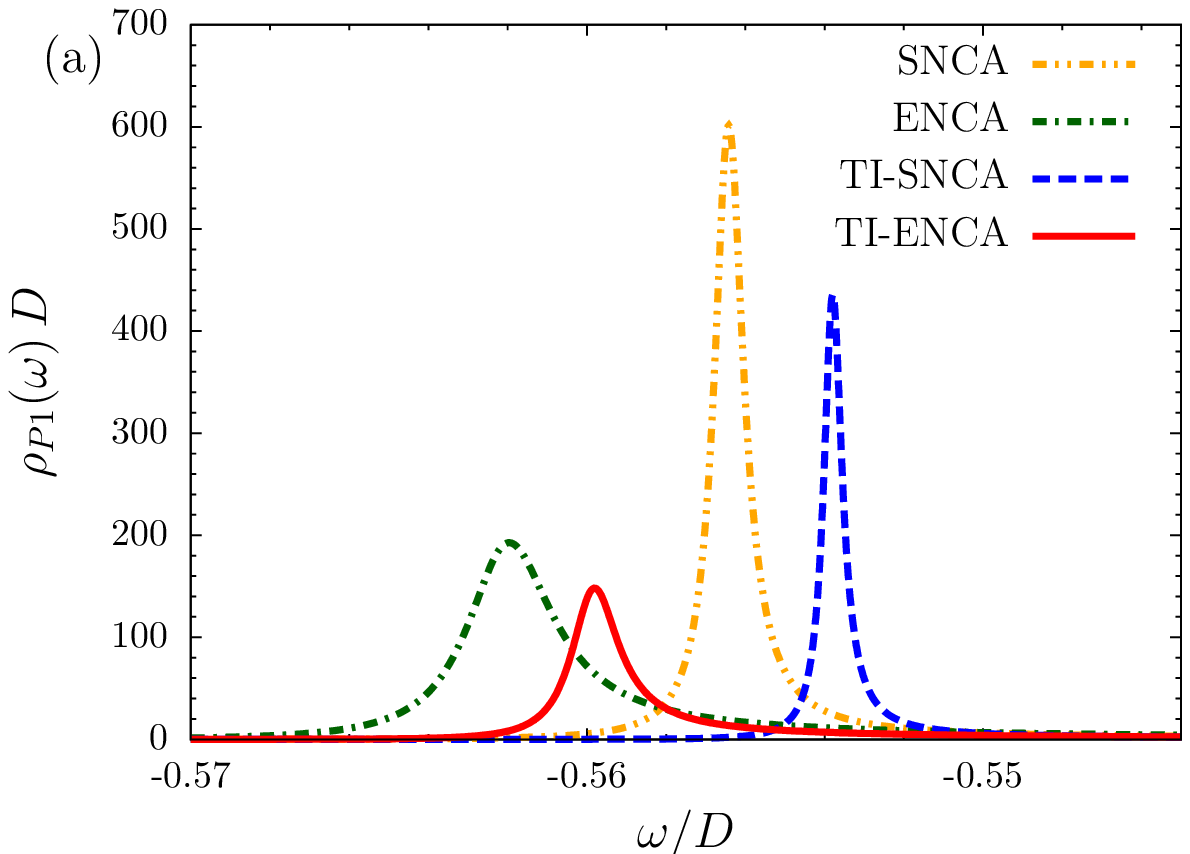}
  \label{fig:3b}\includegraphics[width=0.9\linewidth]{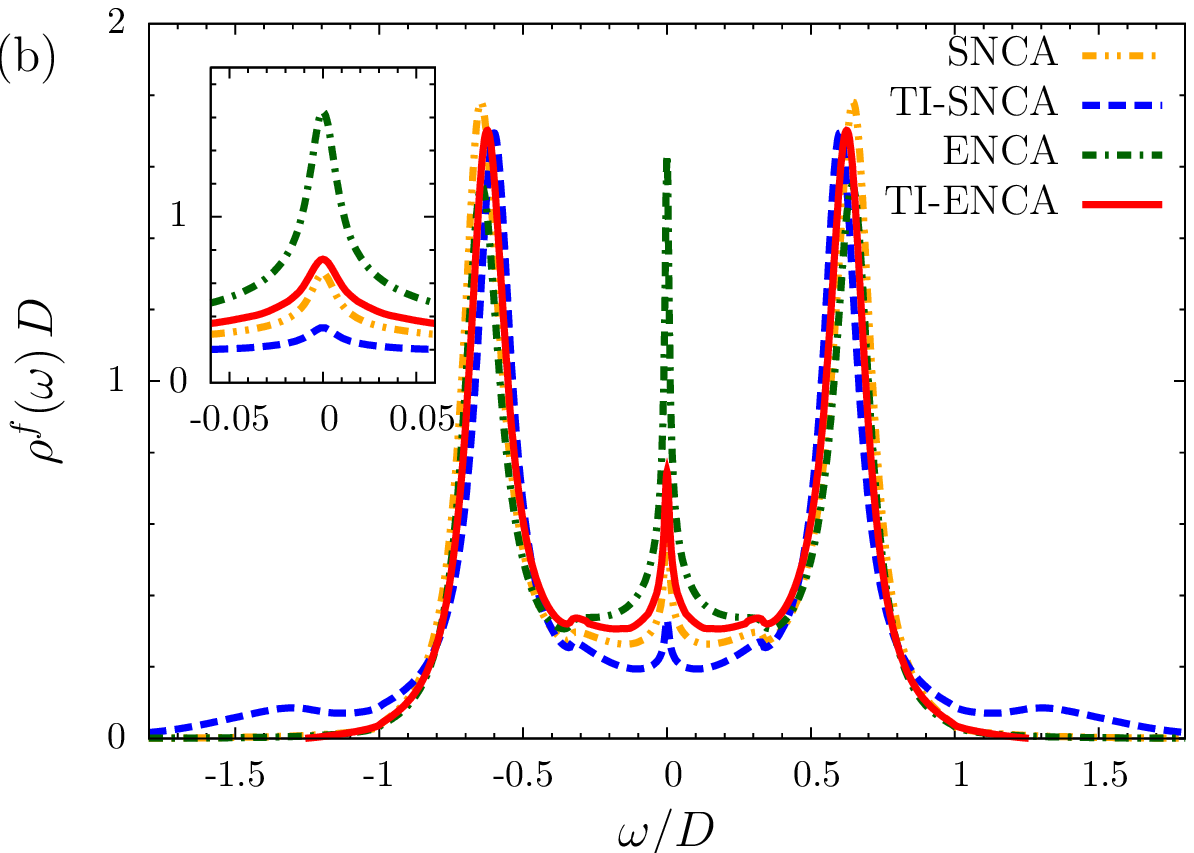}
\caption{(Color online) (a) Threshold behavior of the different SIAM and the
  decoupled TIAM approximations deduced from the ionic propagator for
  one electron per site and for $\eps^f=-\frac D2$, $U =D$,
  $T=0.003D$, and $\Delta= 0.1D$. 
  For the SIAM $\rho_{P1}(\om) =-\frac 1\pi$Im$[P_{\si,\si}(\om + i\delta)]$
  and for the TIAM as
  $\rho_{P1}(\omega)=-\frac 1 \pi$Im$[P_{\sigma\sigma',\sigma\sigma'}(\omega/2+i\delta)]$.
  (b) Full one-particle spectral functions of the different approximations for the same parameters 
  as in
  (a). The inset shows a close-up of the region around the Fermi level.}
\label{fig:res3}
\end{figure}
Calculations for the Kondo regime are presented in Figs.~\ref{fig:res3}(a) and \ref{fig:res3}(b).
Care has to be taken to facilitate a comparison with the results of
single-ion SNCA- and ENCA-solvers, which are shown in these figures,
too. Since our two-impurity SNCA  and ENCA calculations treat the two-impurity
systems, though not coupled, as a whole, the groundstate energy and
threshold energies for local excitations should have twice their single-ion
value, and the single-particle spectral functions are normalized to two.
Hence the energies and  curves are correspondingly scaled with a factor of two to be comparable. 

Figure \ref{fig:res3}(a) shows the spectrum of the  ionic propagator 
in the energy region around the ionic threshold 
for singly occupied impurities, i.e.\ 
$\rho_{P1}(\omega)=-\frac 1 \pi$Im$[P_{\sigma,\sigma}(\omega+i\delta)]$ for the SIAM and 
$\rho_{P1}(\omega)=-\frac 1 \pi$Im$[P_{\sigma\sigma',\sigma\sigma'}(\omega/2+i\delta)]$  for the two-impurity  model.
As is well
known\cite{greweCA108,*schmittSus09} the ENCA performs considerably better than the SNCA
since, for instance, it  produces a more accurate, i.e.\ larger, Kondo temperature $T_\mathrm{K}$.
$T_\mathrm{K}$ can be estimated as the difference  between the threshold, i.e.\ the position of the 
maximum in $\rho_{P1}$, and the single-particle energy, 
$T_\mathrm{K}\sim E_\mathrm{threshold}-\epsilon^f$. 
The same apparently  holds when comparing of the two-impurity SNCA  and two-impurity ENCA 
approximation.
The two-impurity ENCA even seems to exceed 
the single-impurity SNCA and comes near to the quality of the single-impurity
ENCA, at
least for the typical parameter values shown.

A similar conclusion can be drawn from the  one-particle spectral functions
of Eq.~\eqref{eq:rhof}
shown in Fig.~\ref{fig:res3}(b).
They exhibit the typical three-peak
structure with a Kondo-resonance near the Fermi-level and large
spectral weight at the so-called Hubbard peaks which originate from the 
ionic one-particle energies $\eps^f$ and $\eps^f+U$. 
While the high-energy features of the SIAM and TIAM solutions are in good agreement,
the low-energy Kondo resonances of the TIAM are less pronounced.
All 
curves are calculated with the same absolute temperature $T=0.0003D$
which implies different relative temperatures to the corresponding
Kondo scale of each approximation. The TIAM solutions produce a slightly 
smaller Kondo scale and consequently the peaks are not as high, which
confirms the conclusions drawn from the ionic propagators.

The good qualitative and mostly also quantitative agreement
of the calculations support our confidence in the validity
of the results presented for coupled impurities in the next section.  

\section{Results for coupled impurities}
\label{sec:iii}

The physics of a fully coupled cluster of two impurities
can become amazingly rich when all possible sources of interactions
are taken into account. We therefore restrict ourselves to a few
important cases, which we first will consider separately
in order to clarify basic physical effects. 

In general, the hybridization part of the Hamiltonian
will not only induce the tendency towards a
local Kondo-screening of moments considered in the previous section, but will also
induce an additional effective RKKY-exchange interaction with the asymptotic form
in leading order perturbation theory\cite{fazekas:LectureNotes99}
\begin{align}
  \label{eq:9}
  J_\mathrm{RKKY}&\propto -V^4  \frac{\sin(2k_\mathrm{F}d)+2k_\mathrm{F} d 
    \cos(2 k_\mathrm{F} d)}{\left(2k_\mathrm{F}d\right)^4 },
\end{align}
where $k_\mathrm{F}$ is the Fermi wave vector along
the distance  $d\gg1$  between the impurities (lattice constant $a=1$).
The RKKY-interaction according to
Eq.~\eqref{eq:9} oscillates with distance and can induce a ferromagnetic
or  antiferromagnetic coupling.

\subsection{Isolated two-impurity cluster}
\label{sec:atomic}

It is instructive to consider, what physics already is included
without hybridization $\hat V$ to the bandstates.
For this purpose one can inspect the exact solution of the eigenvalue
problem for the isolated cluster $(\hat V=0)$ in the presence of 
interactions
which amounts to setting 
all hybridization functions to zero,  $\Delta_{jl}(\omega)=0$.
We include a direct magnetic exchange term 
and also  allow for a direct single-electron transfer, 
\begin{align}
  \label{eq:5}
  \hat W^f=&\sum_{\si}t ( \fd_{1\si} \f_{2\si} + h.c.)
  - J\vek{\hat S}^f_1 \vek{\hat S}^f_2.
\end{align}

Since the exchange interaction has no effect on the one-particle
sector of the cluster, the one-particle eigenstates correspond to the
(anti-)symmetrized operators given in Eq. (\ref{eq:2}) with
eigenenergies
\begin{align}
  \label{eq:singSpilt}
  E^{(1)}_{\pm}=\eps^f_{\pm}=\eps^f \pm t.  
\end{align}

The two-particle sector contains six states, three of which belong to a
degenerate triplet with one electron on each of the sites. For these
states spin-conserving hopping is ineffective due to the
Pauli-principle, which leads to:
\begin{align}
  \label{eq:6}
  & |S=1,m=\si\rangle=|\si\rangle_1|\si\rangle_2,\\\nonumber
  &|S=1,m=0\rangle=\frac{1}{\sqrt{2}}\Big(|\uparrow\rangle_1 |\downarrow\rangle_2+|\downarrow\rangle_1 |\uparrow\rangle_2\Big)
  \intertext{with eigenenergies}
  \nonumber
  &E_T^{(2)}=2 \eps^f-\frac{J}{4}.
\end{align}

The three remaining singlets in the two-particle subspace are  
\begin{align}
  \label{eq:7}
  &|S=0,m=0\rangle_0=\frac{1}{\sqrt{2}}\Big(|\uparrow,\downarrow\rangle_1 |0\rangle_2-|0\rangle_1 |\uparrow,\downarrow\rangle_2\Big)
  \\\nonumber
  &E^{(2)}_{S_0}=2\eps^f+U, 
  \intertext{and}
  \label{eq:7a}
  &|S=0,m=0\rangle_\pm=\frac{1}{N_\pm}\Bigg[\frac{1}{\sqrt{2}}\Big(|\uparrow\rangle_1 |\downarrow\rangle_2-|\downarrow\rangle_1 |\uparrow\rangle_2\Big)
  \nonumber  \\\nonumber
&\phantom{MMMMMMMM}+ a_{\pm}\Big(|\uparrow,\downarrow\rangle_1 |0\rangle_2+|0\rangle_1 |\uparrow,\downarrow\rangle_2\Big)\Bigg]
\nonumber\\
&E^{(2)}_{S_\pm}=2 \eps^f+\frac{3 J}{4}-a_\pm 
\end{align}
where
\begin{align*}
a_{\pm}=&\frac{3J}{8}  - \frac{U}{2} \Big(1 \pm \sqrt{1+\frac{16t^2}{U^2}+\frac{9J^2}{16U^2}- \frac{3J}{2U}}\Big),\\
N_{\pm}=&\sqrt{1+2\frac{t^2}{a_{\pm}^2}}.
\end{align*}

In order to understand the result it is useful to realize 
that for the polar state $|S=0,m=0\rangle_0$  direct transfer is irrelevant due
to orbital antisymmetry. The 
effect of direct hopping  on the remaining two singlets, as
well as the influence of  direct exchange $J$ are most clearly
seen in the limiting 
case where  $U$ exceeds $t$
and $J$,  i.e.\ $U\gg |t|,|J|$, where one finds
\begin{align}
  \label{eq:8}
  & E^{(2)}_{S_+}\approx 2 \eps^f+U+\frac{4t^2}{U},\\
  \label{eq:8b}
  & E^{(2)}_{S_-}\approx 2 \eps^f+\frac{3 J}{4}-\frac{4t^2}{U}.
\end{align}
In this limit the well known competition between the triplet
of Eq.~(\ref{eq:6}) and the singlet of Eq.~(\ref{eq:8b}) for
becoming the ground state is realized. 
The physics is determined by the exchange splitting of these levels,
\begin{align}
  \label{eq:Jt}
  \Delta E=E^{(2)}_{S_-}-E^{(2)}_{T}=J - \frac{4t^2}{U}\equiv J+J_t
  .
\end{align}

For larger values of the direct transfer $t$ the level-splitting
becomes increasingly important and renders a discussion based exclusively
on an effective exchange interaction impossible. 

States with three or four electrons in the cluster have higher
energies through additional contributions of $U$. They can be deduced from
the above  results by particle-hole transformation.

\subsection{Direct exchange coupling}
\label{sec:J}

For the following numerical study we will set $t=0$
but consider a direct coupling $J$. 
We also
include a finite diagonal hybridization but still ignore  the non-diagonal hybridization 
functions producing the RKKY interaction, i.e.\ $ \Delta_{12}(\omega)=0$.
Therefore, we choose a direct two-impurity interaction Hamiltonian [see Eq.~\eqref{eq:Wf}]
\begin{align}
  \label{eq:WJ}
  \hat W^f=& - J\vek{\hat S}^f_1 \vek{\hat S}^f_2
  .
\end{align}

Then,  the typical competition between the single-ion Kondo effects 
and the fixed nonlocal exchange $J$ results. 
The Kondo coupling between each impurity spin and the conduction electrons 
is governed by the antiferromagnetic exchange
\begin{align}
  J_\mathrm{Kondo}=-\frac{2U\Delta}{\eps^f(\eps^f+U)}>0
\end{align}
leading to a characteristic energy scale
\begin{align}
  \label{eq:TK}
  T_\mathrm{K}=\alpha \sqrt{J_\mathrm{Kondo}}\exp\big(-\pi/J_\mathrm{Kondo}\big)
\end{align}
with  $\alpha\approx 2\pi \min\{U,D\}$.

\begin{figure}[ht]
  \includegraphics[width=0.9\linewidth]{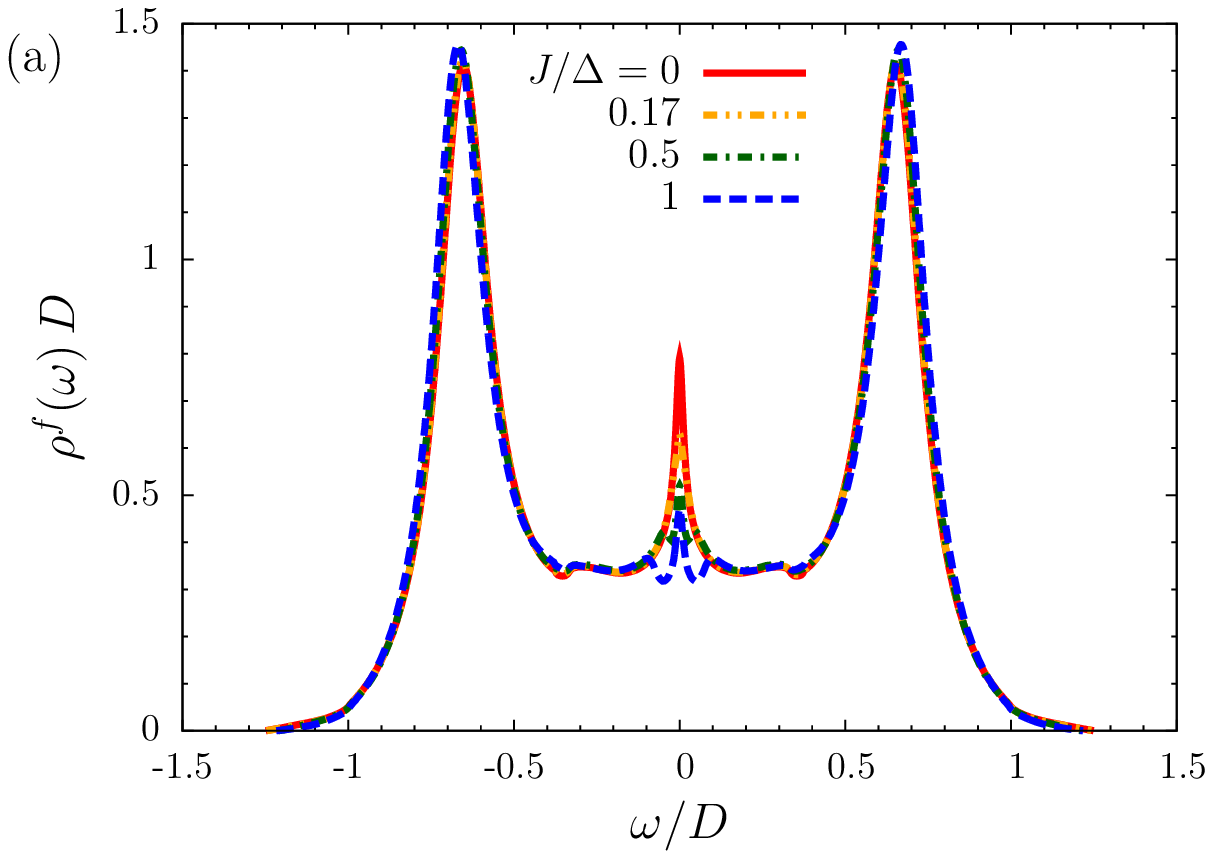}
  \includegraphics[width=0.9\linewidth]{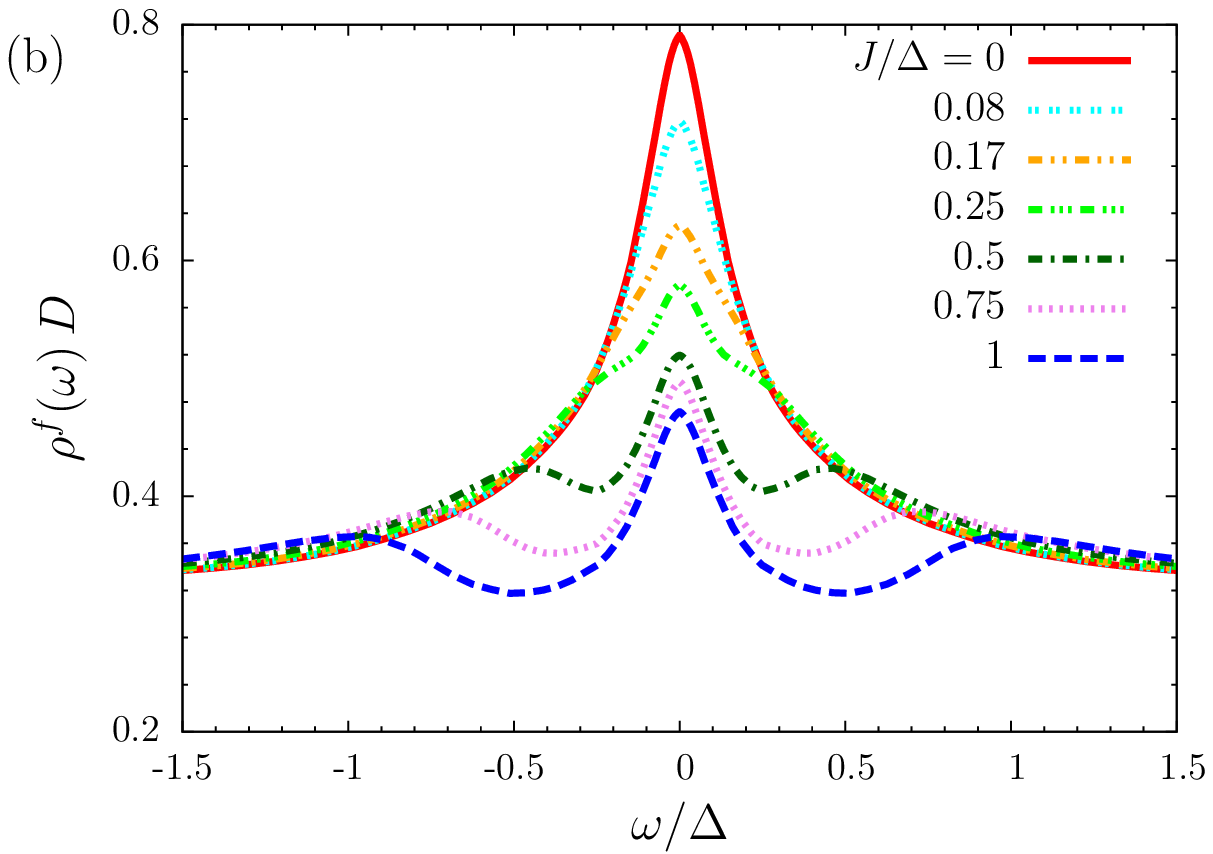}
  \includegraphics[width=0.9\linewidth]{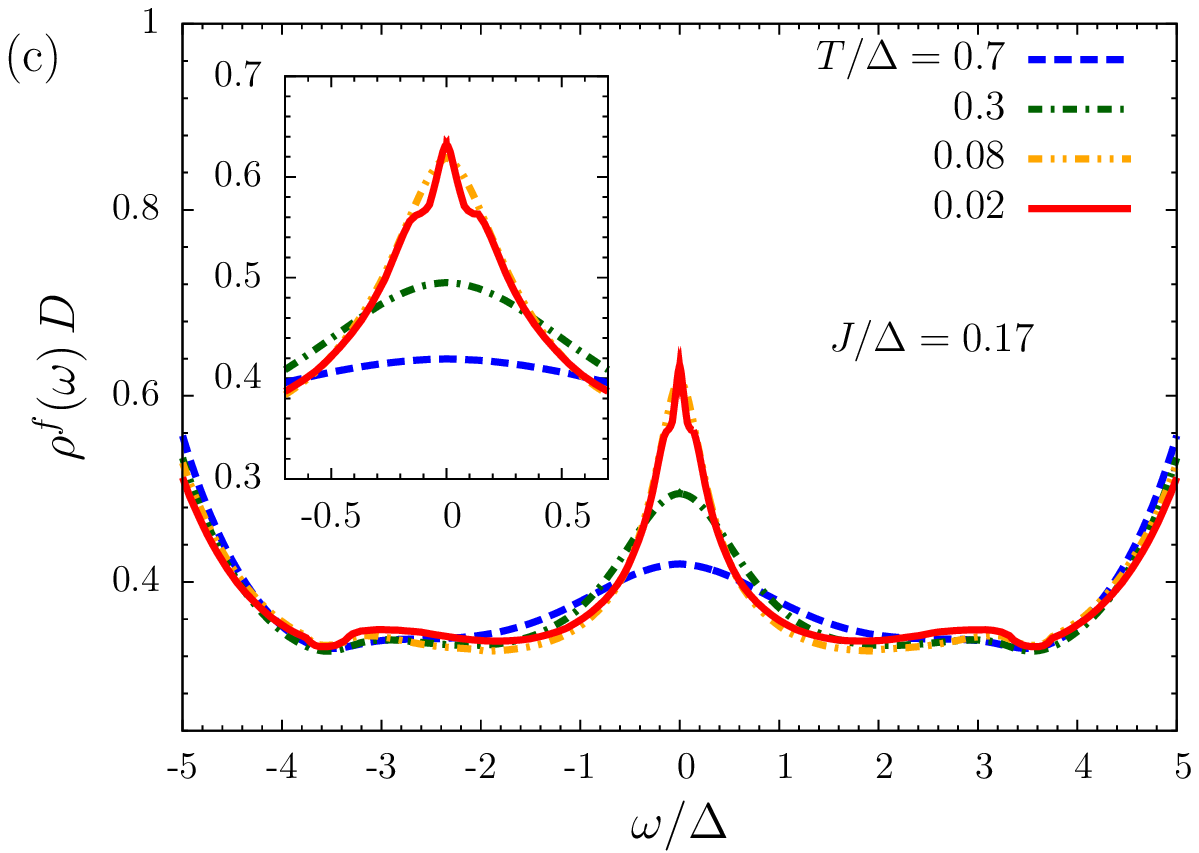}
  \caption{(Color online)  Spectral function for $\epsilon^f=-\frac D2$, $U=D$, 
    $\Delta=0.1D$ with  direct ferromagnetic 
    coupling $J>0$: (a) Fixed temperature $T=0.004D=0.04\Delta$
    and various values of the coupling. (b) Magnification of 
    the region around the Fermi-energy. (c) The spectral 
    function near the Fermi level
    for fixed coupling $J=0.017D=0.17\Delta$ and various temperatures.
    Notice, that while in panel (a) the energy is measured in units of the 
    half-bandwidth $D$, panels (b)  and (c) use the Anderson width $\Delta$.
  }
\label{fig:res4}
\end{figure}
One-particle spectral functions calculated with the two-impurity ENCA-solver 
are shown in Fig.~\ref{fig:res4} for 
various values of a ferromagnetic  direct coupling ($J>0$)
and temperatures. The overall form of the spectral function 
and the high-energy features are essentially unchanged under the inclusion of $J$. 
However, pronounced changes occur in the
low-energy region around the Fermi energy
$\om=0$, which at $J=0$ and small $T\lesssim T_\mathrm{K}$
exhibits the well-known many-body  resonance of the 
single-ion Kondo effect.

For very large direct exchange coupling $J>0$ the
two impurity spins are aligned parallel and nearly act as one rigid
spin with  $S=1$. This corresponds to the triplet state in the isolated cluster
which experiences a modified two-stage nonlocal Kondo screening
with a reduced characteristic energy scale $T_{\mathrm{K}}^{\mathrm{(2imp)}}<T_\mathrm{K}$.\cite{silvaTwoImpNRG96}
This leads to the formation of a narrower and less saturated Kondo resonance 
at $\om=0$, clearly visible in Fig.~\ref{fig:res4}(b)  and
is similar to what happens in  multi-orbital
SIAM.\cite{kuboHundSIAM99,*pruschkeTwoBandHMNRG05,*nevidomskyyHundJKondo09} 

At excitation energies of the order of 
the exchange coupling, $\omega \approx \pm J$, one probes a domain 
dominated by correlations typical for the single-ion Kondo effect.
The weaker nonlocal correlations  of the low-energy regime are then broken. 
The two  impurities appear as essentially uncoupled and
behave like isolated single impurities. In the spectral functions this 
is indicated by the humps visible  for small $J$ which
develop into side-maxima at larger $J$  which are located the energies 
$\omega\approx \pm J$. For energies larger than $|\omega|> J$
the $J=0$ spectral function is approached. This is in accord with
the finding, that the characteristic high-energy scale 
of this model is essentially given by the single-impurity
Kondo scale.\cite{zhuTIAM11}

This is also reflected in the temperature dependency as shown in
Fig.~\ref{fig:res4}(c). The many-body resonance at first forms 
like in a SIAM, and only for low temperatures $T\lesssim J$ the narrowing 
near the Fermi level sets in and causes the above 
mentioned humps or maxima.

\begin{figure}[ht]
  \centering
  \includegraphics[width=0.9\linewidth]{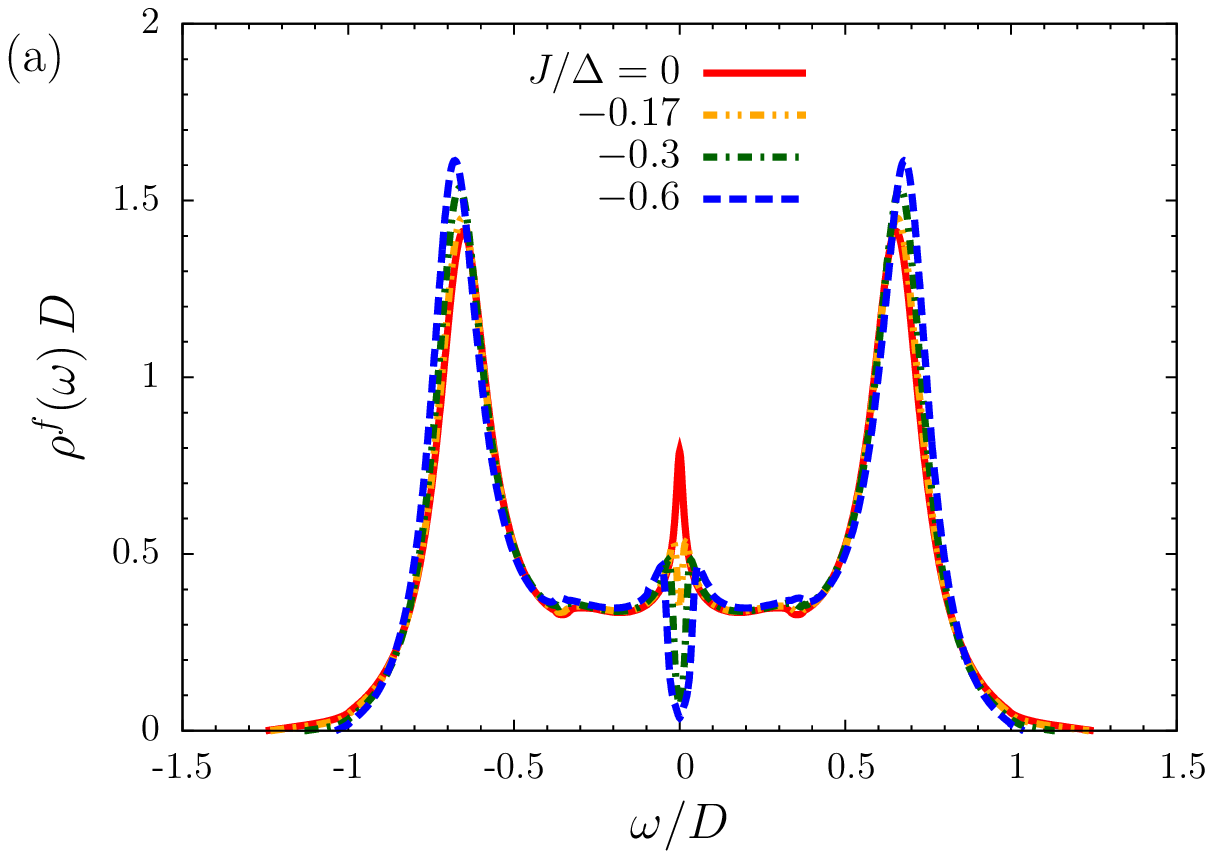}
  \includegraphics[width=0.9\linewidth]{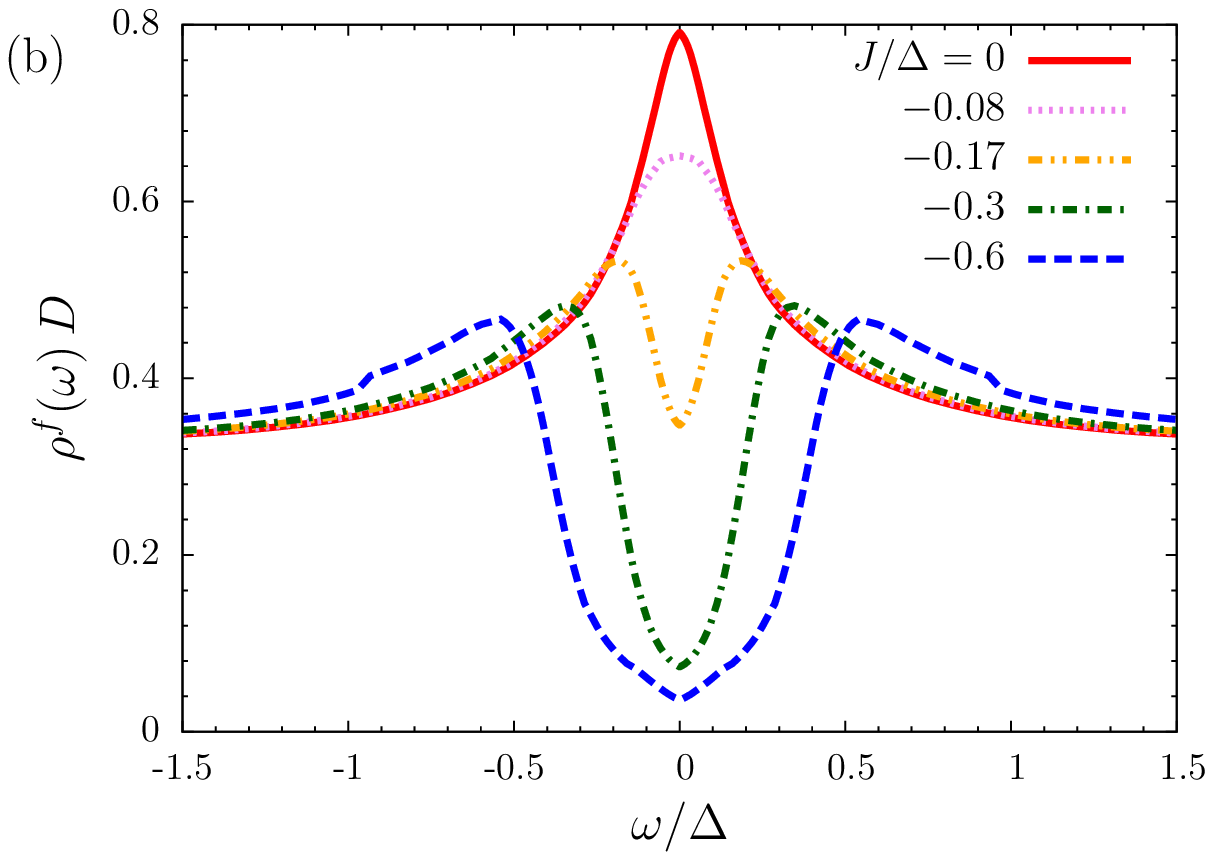}
  \includegraphics[width=0.9\linewidth]{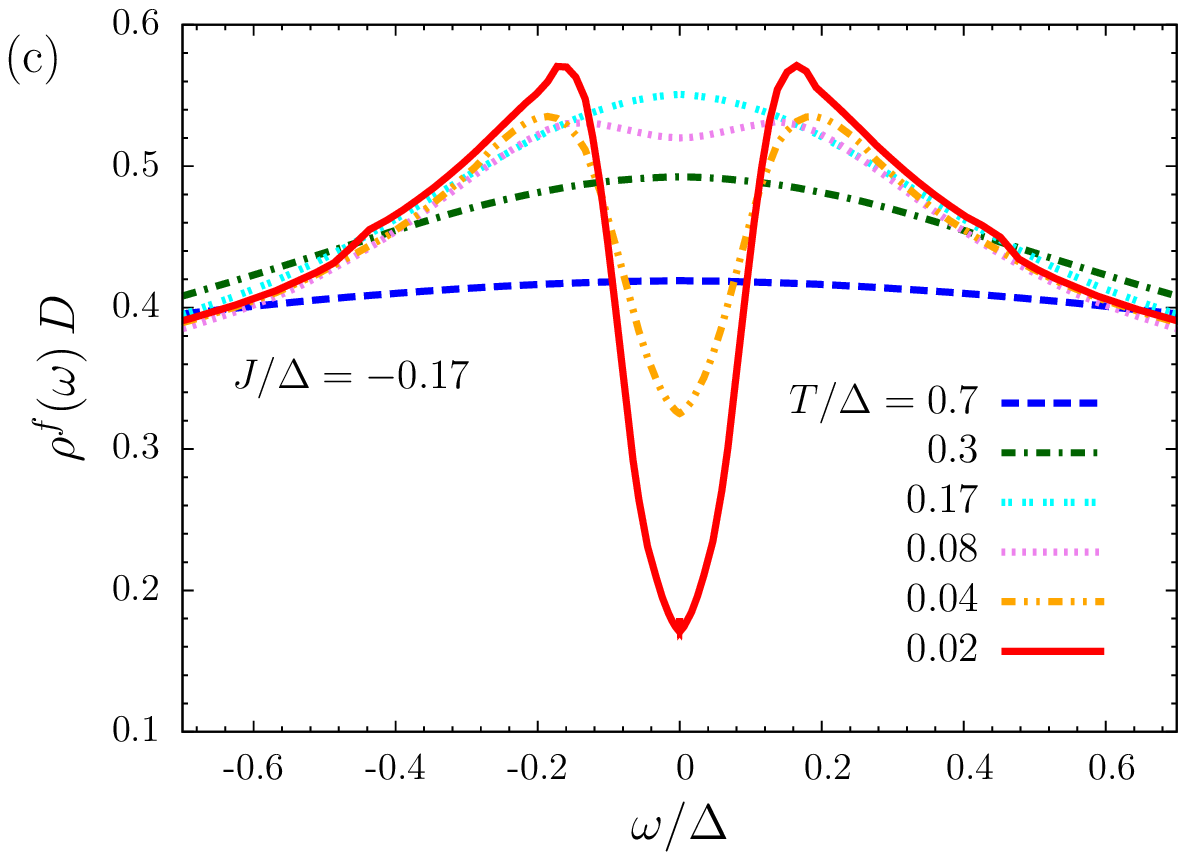}
  \caption{(Color online)  (a) Spectral functions for various antiferromagnetic direct exchange couplings $J<0$
    and fixed temperature $T=0.04\Delta$. Panel (b) shows a magnification of the
    low energy region. 
    (c)  The spectral function around the Fermi level
    for fixed coupling $J=-0.17\Delta$ and various temperatures.
    Notice, that while in panel (a) the energy is measured in units of the 
    half-bandwidth $D$, panels (b)  and (c) use the Anderson width $\Delta$.
    Other parameters are as in Fig.~\ref{fig:res4}.
  }
\label{fig:res5}
\end{figure}
We now turn to the case of an antiferromagnetic direct coupling 
($J<0$) where  the nature of the physical scenario for low
lying states changes qualitatively. 
Now  a real competition takes place between local singlet formation via the
Kondo-effect and nonlocal singlet binding in the molecular
two-electron state. The single-ion Kondo effect present at small 
$J$ is suppressed with increasing negative values of $J$.
This is clearly borne out by the spectral functions of Fig.~\ref{fig:res5}, which
exhibit a very rapid depletion of the spectral function near
the Fermi energy  with $|J|$. 

The formation of a gap around $\om=0$ then implies that the low lying 
molecular singlet does effectively not interact with the band states.
However, side peaks at the edges of the gap can be observed
at energies $\omega=\pm J$.
These are indicative for the fact that
the excited molecular triplet experiences a kind of virtual Kondo
effect.\cite{paaskeNeqKondo06}

This interpretation is supported by the temperature dependence
of the curves depicted in Fig.~\ref{fig:res5}(c) for a fixed value of $J=-0.17\Delta$.
In the same way as the pseudogap around $\om=0$ forms 
for temperatures lower than $|J|$, the side peaks emerge and increase in
height.

\subsection{Direct single-particle hopping}
\label{sec:hopping}

We now neglect the direct magnetic exchange, 
but introduce a direct single-particle hopping, i.e.\
\begin{align}
  \label{eq:Wt}
  \hat W^f=&t\sum_{\si} ( \fd_{1\si} \f_{2\si} + h.c.).
  \intertext{We still ignore the non-diagonal hybridization}
  \Delta_{12}(\omega)&=0
  .
\end{align}

Based on the  discussion of the isolated two-impurity cluster in 
Sect.~\ref{sec:atomic}
we can expect a twofold source for modifications of 
a pure single-ion Kondo effect:
(1) $t$ gives rise to an effective antiferromagnetic exchange
interaction $J_t=-\frac{4t^2}{U}$, and (2) $t$ splits the ionic one-particle levels
$\eps^f \rightarrow \eps^{f}_{\pm}$ and produces a tendency towards even
and odd molecular one-particle states. 
Therefore, the interesting
question arises, how the scenario developed for an antiferromagnetic
direct exchange interaction will be altered by the impending effect of
even-odd splitting.

\begin{figure}[ht]
  \centering
  \includegraphics[width=0.9\linewidth]{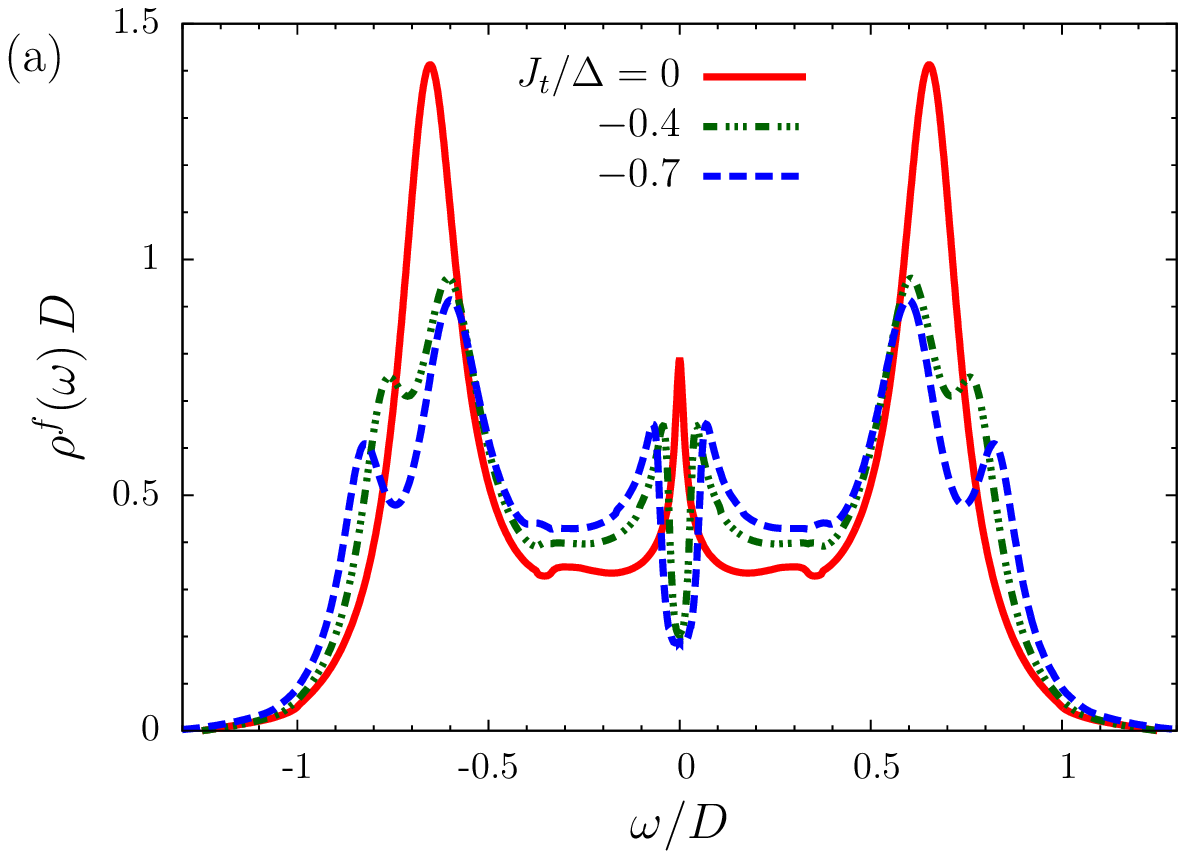}
  \includegraphics[width=0.9\linewidth]{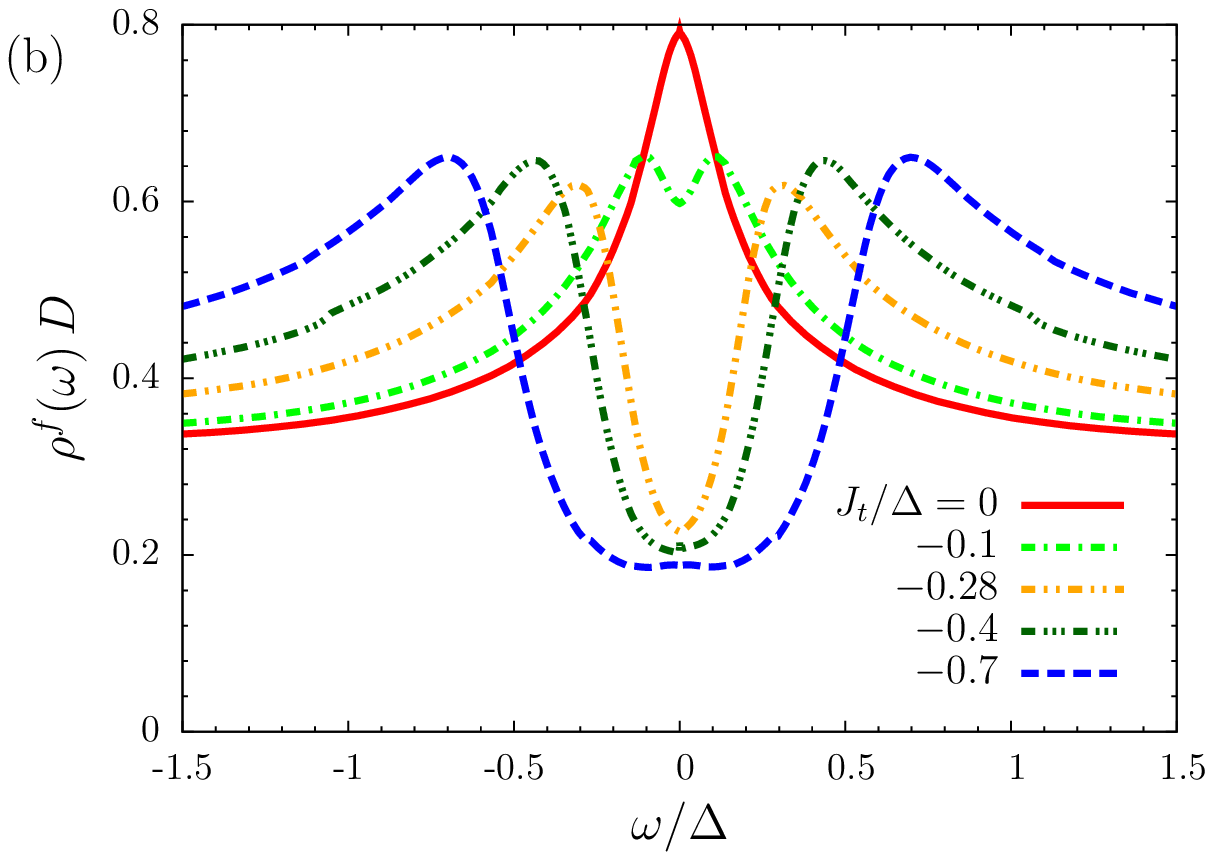}
  \caption{(Color online)  (a) Spectral functions  at a fixed temperature $T=0.04\Delta$ for various values 
    of the direct single-particle hopping $t$. To reveal its physical influence more clearly
    the effective antiferromagnetic exchange $J_t=-\frac{4t^2}{U}$ is given instead
    of the bare hopping $t$. For example, $J_t=-0.4\Delta$ corresponds to $t=0.1D$
    and  $J_t=-0.7\Delta$ corresponds to $t=0.13D$.
    Panel (b) shows a magnification of the low energy region (energy is measured in 
    units of the Anderson width $\Delta$).
    Other parameters are as in Fig.~\ref{fig:res4}.
  }
\label{fig:res6}
\end{figure}

Figure \ref{fig:res6} shows the spectral function for various values of $t$. 
One recognizes the expected opening of the
pseudogap around the Fermi level $\om=0$ with increasing absolute value of
$J_t$ [see panel (b)]. 

The hopping induces changes of the spectral function at all energies
[see \ref{fig:res6}(a)], 
which is in contrast  to the case of a direct coupling where only the 
low-energy region was affected.  Particularly, and in accord with our 
expectations, the ionic Hubbard resonances near $\om=\eps^f$ and 
$\om=2\eps^f+U$ are split by the hopping. 
The separation of the peak maxima 
is roughly given by $2t$ which is in accord with the simple reasoning 
from the isolated two-impurity 
model, see Eq.~\eqref{eq:singSpilt}.
This reflects  a tendency, which has been outlined before\cite{grewe:quasipart05},
namely that features of the unperturbed structures of the one-particle states leave their
traces in the quasiparticle bandstructure. 

In the low-energy region, 
hopping  produces a gap similar to the case of a direct exchange coupling
[see Figs.~\ref{fig:res5}(b) and \ref{fig:res6}(b)].
However, whereas the gap is not as 
pronounced as in Fig.~\ref{fig:res5}(b),
the side peaks at the edge of the gap are considerably 
higher. 

\begin{figure}[ht]
  \centering
  \includegraphics[width=0.9\linewidth]{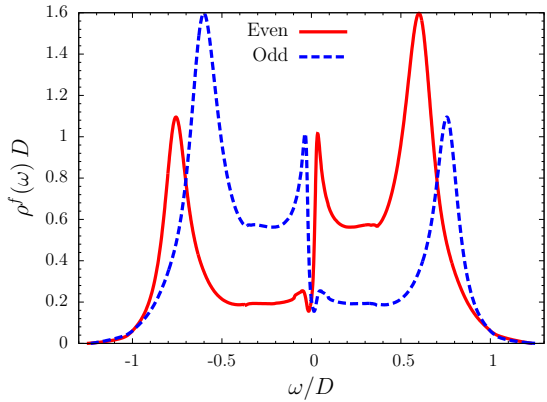}
  \caption{(Color online)  Spectral function of Fig.~\ref{fig:res6}(a) for $J_t/\Delta=-0.4$
    decomposed onto even and odd contributions. 
  }
  \label{fig:res8}
\end{figure}

The question arises whether the splitting is due to a suppression of the 
Kondo effect as in the case of a direct antiferromagnetic $J$.  It can be 
addressed by decomposing the spectral function into even and odd parity parts
as would be produced by the operators $\fd_{\pm \si}$ of 
Eq.~(\ref{eq:2}). This is shown in Fig.~\ref{fig:res8}
for $t=0.1D$, i.e.\ $J_t=-0.4\Delta$. 
All the peaks in the spectral function can clearly be attributed to  
(mostly) one parity channel. In particular,
the quasiparticle Kondo resonance is  partitioned
into an odd contribution for negative energies 
and an even contribution for positive energies.   

This suggests that the splitting is not exclusively due to a suppressed 
Kondo effect, but is mainly a single-particle  even-odd splitting of the 
quasiparticle excitations.
The side peaks then  indicate the local  preformation of quasiparticle bands, with the odd 
states associated with the bonding region at $k=0$ and the even states with the
anti-bonding region at $k=\pm \pi$. 

\begin{figure}[ht]
  \centering
  \includegraphics[width=0.5\linewidth]{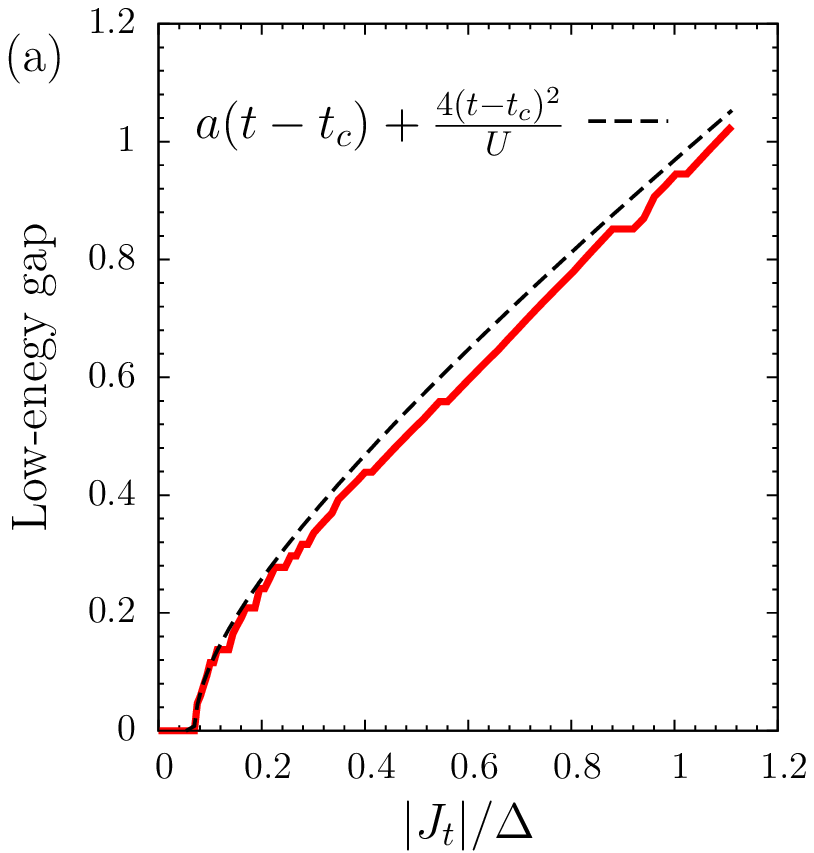}%
  \includegraphics[width=0.5\linewidth]{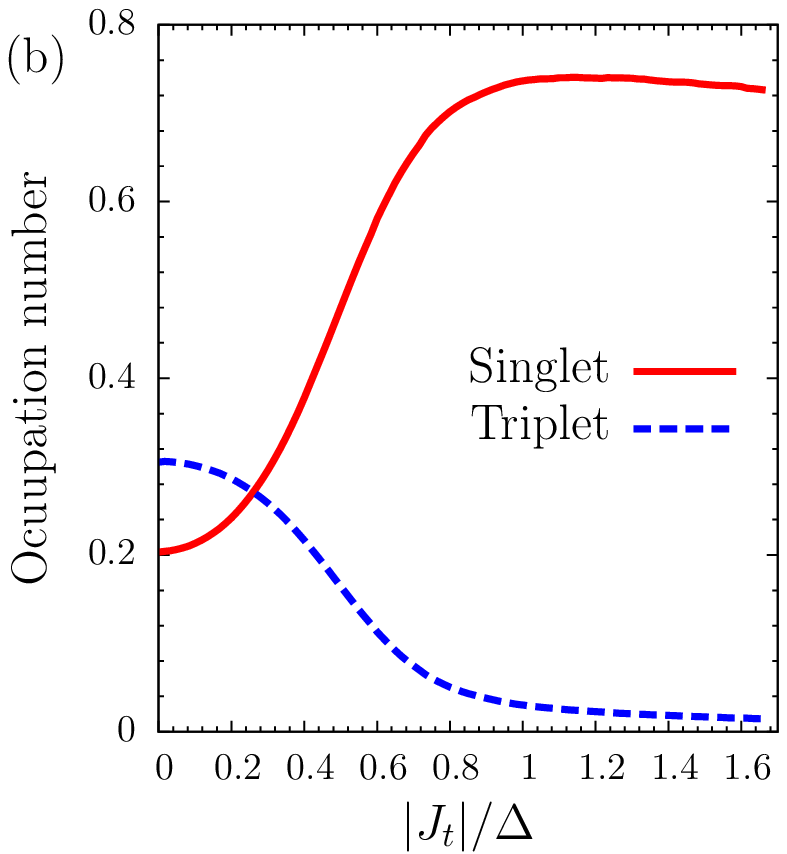}
  \caption{(Color online)  (a) Low-energy gap as function of $|J_t|$. 
    The parameters for the dashed fit are $t_c=0.05\Delta$ and $a= 2.16\Delta$   
    (b) Occupation numbers of the singlet and triplet states 
    in the two-particle sector as function of the (negative) effective 
    antiferromagnetic exchange  $J_t$.
    All other parameters are as in Fig.~\ref{fig:res6}.
  }
  \label{fig:res7}
\end{figure}

We show the width of the central low-energy pseudogap 
as function of the effective exchange $|J_t|=\frac{4t^2}{U}$ in Fig.~\ref{fig:res7}(a).
The gap first opens at a finite value $t_c$ for  the hopping.   
For smaller values of $J_t$ the splitting behaves like a square root
indicating  an opening  of the gap which is linear in $t$.
At larger $t$ the gap is  dominated by the impact of the quadratic
$t$-dependence of the 
exchange coupling and thus goes linear in $|J_t|$.
Indeed a fit with a function $a(t-t_c)+4\frac{(t-t_c)^2}{U}$ 
reproduces the gap quite well. 

The molecular triplet and singlet states are separated in energy 
roughly  by $|J_t|$ [see Eq.~\eqref{eq:Jt}].
At large $t$ the singlet becomes lower in energy that the triplet. This 
can be directly observed in Fig.~\ref{fig:res7}(b), where
the  occupation numbers of these molecular states are shown as function
of $|J_t|$.  For small  $|J_t|$ the occupation numbers are roughly 
equal and mixed at a nearly constant ratio.
With increasing coupling, the singlet occupation rises steeply,
accompanied by a corresponding  fast depletion of the triplet
states.
The large-$|J_t|$ regime thus reproduces the behavior 
already known from a direct antiferromagnetic exchange.

\subsection{Non-diagonal hybridization and RKKY exchange}
\label{sec:rkky}

\begin{figure}[ht]
  \centering
  \includegraphics[width=0.9\linewidth]{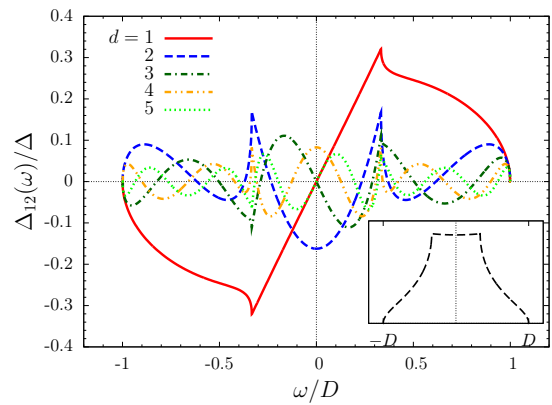}
  \caption{(Color online)  Non-diagonal hybridization  function of Eq.~\eqref{eq:10}
    for various distances as indicated. The inset sketches the
    three-dimensional free density of states of the conduction 
    electrons.  
   }
  \label{fig:res9}
\end{figure}

It is left now to clarify the role played by the non-diagonal
hybridization. We set the direct parameters $t$ and $J$
to zero and include the full hybridization function of Eq.~\eqref{eq:32},
which allows for a coherent hybridization of the cluster states
with the conduction band. The non-diagonal component 
for distance $\vek{d}=\vek{R}_2-\vek{R}_1$ reads 
\begin{align}
  \label{eq:10}
  \Delta_{12}(\om)=& \frac 1N \sum_{\vk}V^2\cos (\vk\vek{d})\pi\delta(\om-\eps^c_{\vk})
  .
\end{align}
The process of non-diagonal hybridization is qualitatively different from a direct hopping $t$
as considered before. As a consequence of the intermediate propagation through the 
band, a nontrivial dependence on the excitation energy $\omega$ results. 
The hybridization function exhibits an increasingly  oscillatory behavior with growing distance between the
impurities  while the absolute height decreases as shown in Fig.~\ref{fig:res9}.
 
The following effects are expected as consequences of diagonal and 
this non-diagonal hybridization: (1) the RKKY-exchange
interaction between the two impurity spins is induced, which favors
ferro- or antiferromagnetic alignment depending on the distance. (2)
Kondo screening-clouds (see, for example, Ref.~\onlinecite{bordaKondoCloud07,*buesserKondoCloud10,*mitchellKondoCloud11,*prueserKondo11})
for singly occupied impurities are induced in the band.
In case the impurities are spatially not well separated and the individual Kondo-clouds
penetrate each other, this then involves considerable  nonlocal coherence.
(3) A tendency toward the stabilization of even and odd parts of
the molecular single-electron orbitals might appear, as 
the result of an induced (indirect) transfer via the band states. 

Thus, we do  not expect a simple realization of the Doniach-scenario 
of competing local and
antiferromagnetic correlations, which would either lead
to a Fermi liquid with quenched spins (and some short-ranged
antiferromagnetic correlations) or to  antiferromagnetic
order between the impurities.

\begin{figure}[ht]
  \centering
  \includegraphics[width=0.9\linewidth]{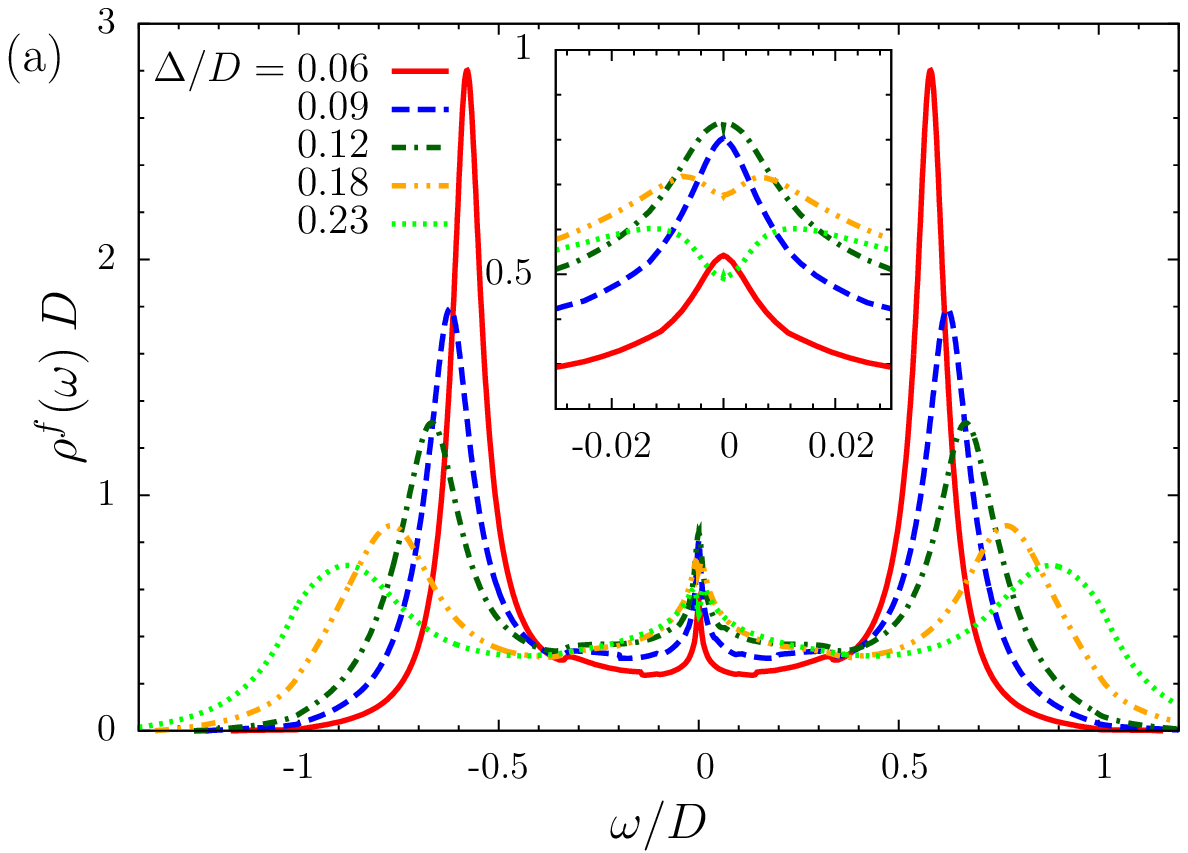}
  \includegraphics[width=0.9\linewidth]{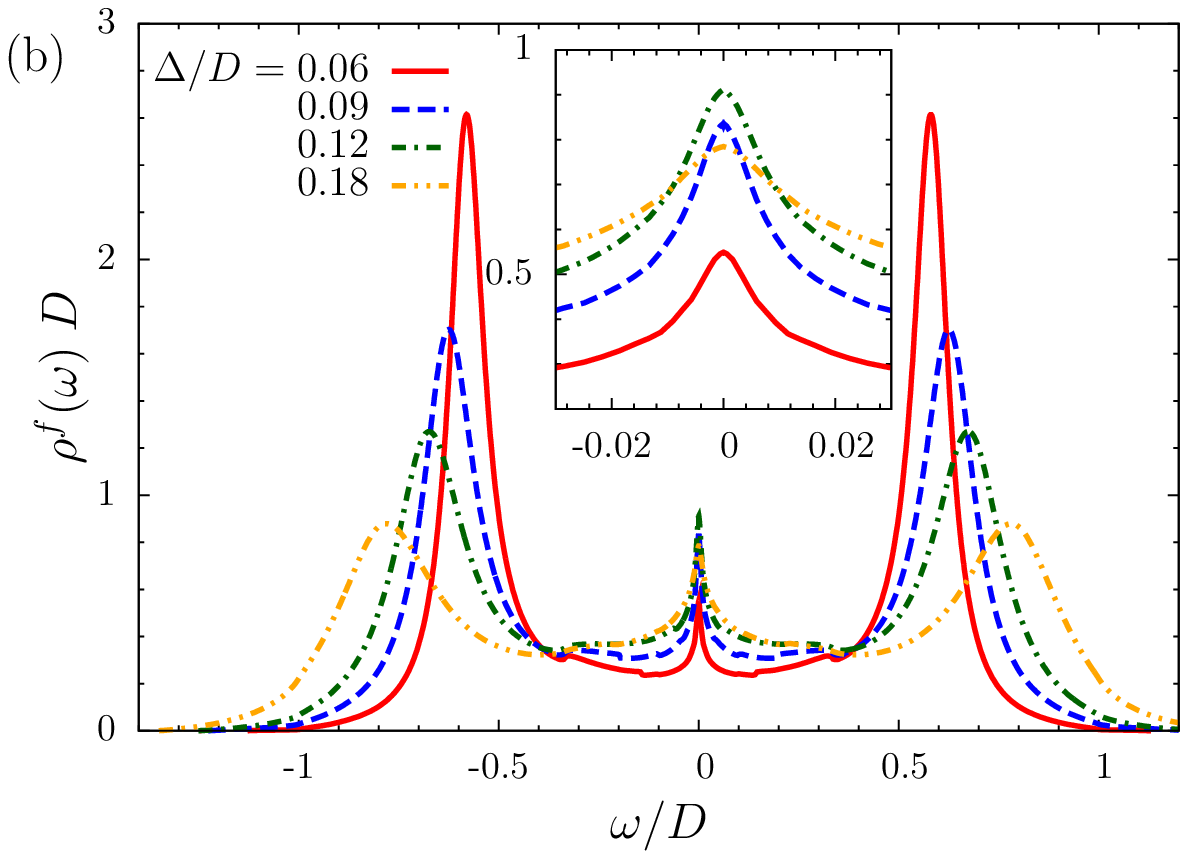}
  \caption{(Color online)  Impurity spectral functions with diagonal and non-diagonal 
    hybridization functions for two distances (a) $|\vek d|=1$
    and (b) $|\vek d|=2$ for various Anderson widths as indicated.
    The inset shows a close-up of the low energy region. 
    Other parameters are: $\epsilon^f=-\frac D2$, $U=D$, 
    $T=0.004D$ and $J=t=0$.
  }
  \label{fig:res10}
\end{figure}

In Figure \ref{fig:res10} the one-particle impurity 
spectral functions for two different
impurity distances along the same principal direction of a three-dimensional simple-cubic
lattice are shown  for
various  values of the hybridization strength $\Delta$.
These cases can loosely be associated with situations
already studied above.  Our choice of parameters places the
nearest-neighbor $d=1$ case into a regime of antiferromagnetic
exchange, $J_\mathrm{RKKY}<0$, whereas at $d=2$ the impurities 
should experience a ferromagnetic $J_\mathrm{RKKY}>0$. 

The spectra greatly  change with increasing hybridization. The 
Hubbard peaks are broadened and are moved to larger energies, as it is expected. 
These high-energy features are mainly determined by the diagonal part of 
the hybridization and are insensitive to the distance between 
the impurities, thus the figures of panel (a) and (b) are very similar 
at large $|\omega|$.
Increasing $\Delta$ first implies a larger absolute value
of $J_\mathrm{RKKY}<0$ for $d=1$. Consequently,  a pseudogap forms at the Fermi level
for $\Delta\gtrsim 0.15 D$ as visible in the inset of Fig.~\ref{fig:res10}(a).
The reason for the gap being  not visible at smaller $\Delta$ is the finite 
temperature of $T=0.004D$ used in the calculation.

For distance $d=2$ the larger ferromagnetic $J_\mathrm{RKKY}$ 
should lead to a narrowing of the many-body resonance with  
side-peaks, but this is not clearly observed in Fig.~\ref{fig:res10}(b).
On the other hand, the local single-impurity Kondo scale is also enhanced with 
increasing $\Delta$ [see Eq.~\ref{eq:TK}],
leading to an increase in the overall spectral weight of the Kondo resonance at the Fermi 
level. (This is also observed  in the case of the antiferromagnetic $d=1$ situation.)
The signatures of indirect ferromagnetic exchange are by 
far not as clear as in the case studied in a previous section with direct exchange.
For one, they are strongly competing with the increasing single-ion Kondo effect
and for another, the dynamic non-diagonal exchange does not act in the 
same simple way as an static exchange
constant.
Therefore, the Doniach scenario seems to be too oversimplifying.

\begin{figure}[ht] \centering
  \includegraphics[width=0.5\linewidth]{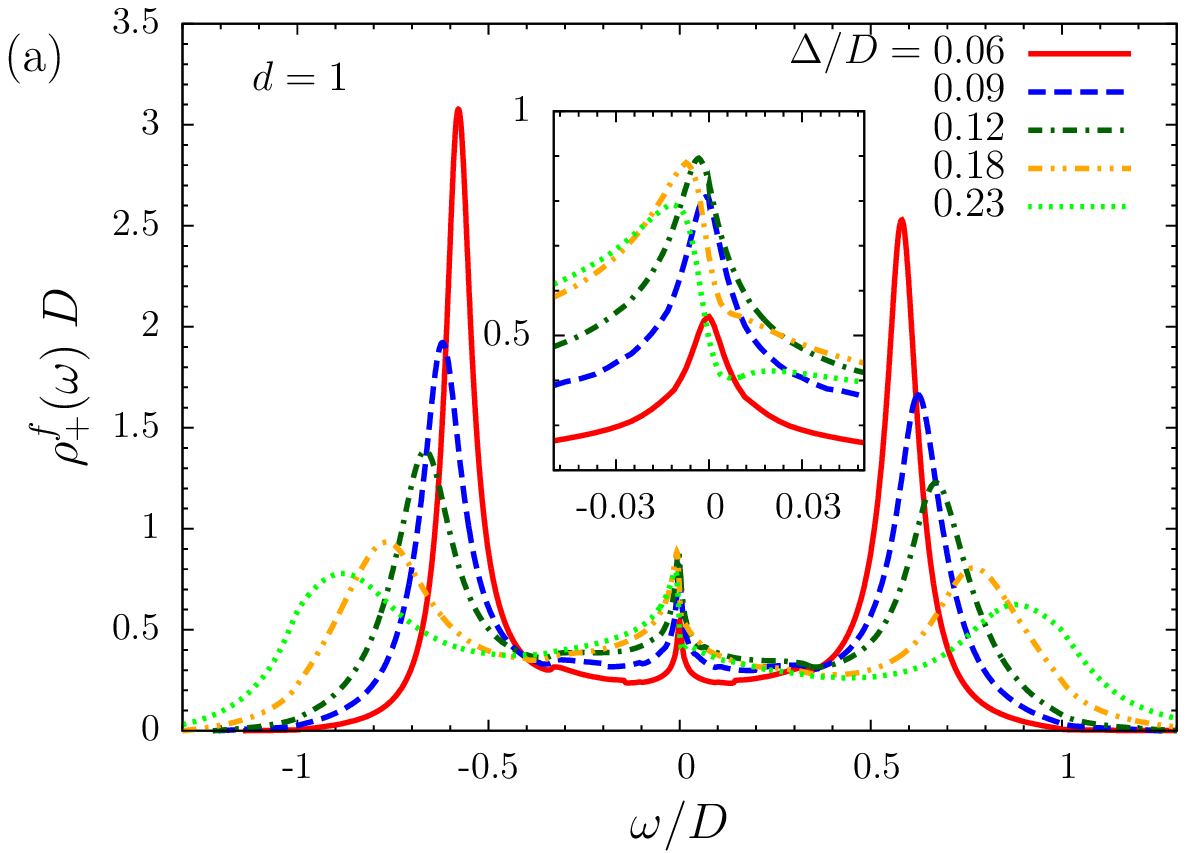}%
  \includegraphics[width=0.5\linewidth]{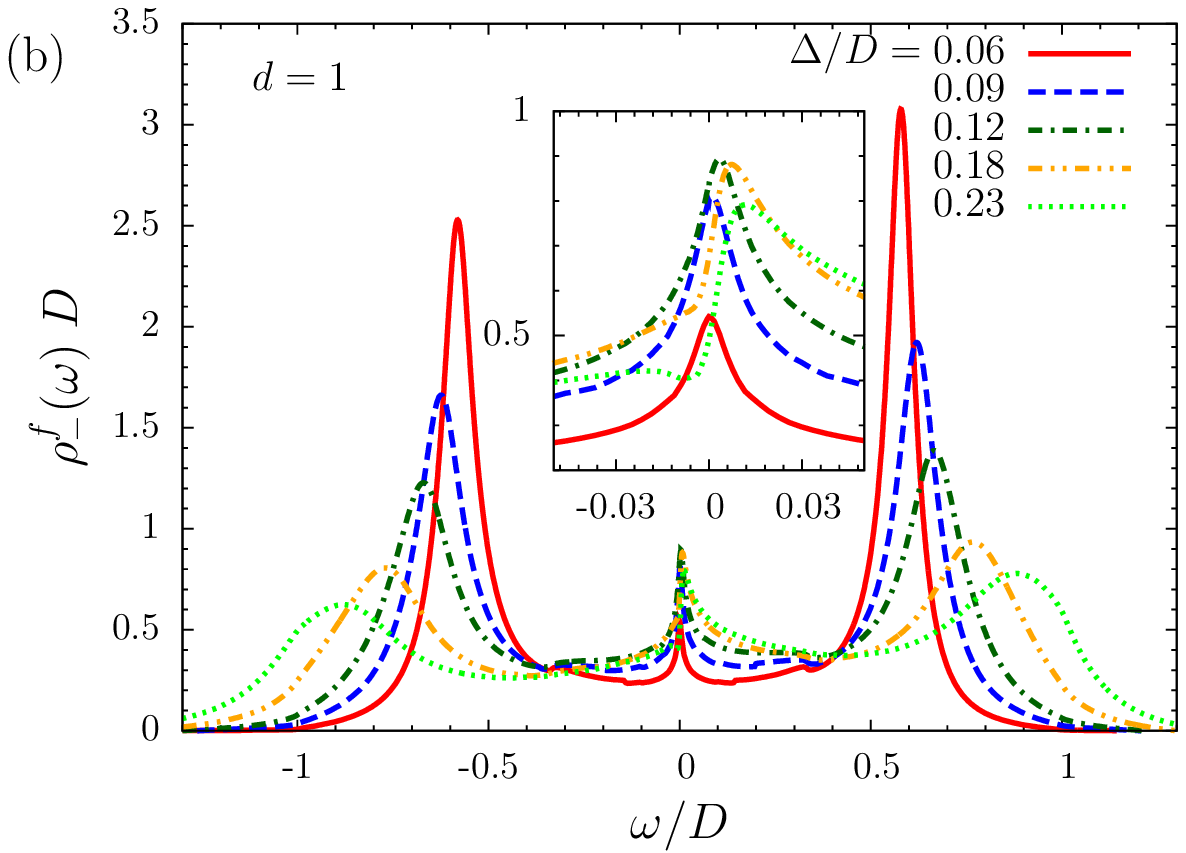}

  \includegraphics[width=0.5\linewidth]{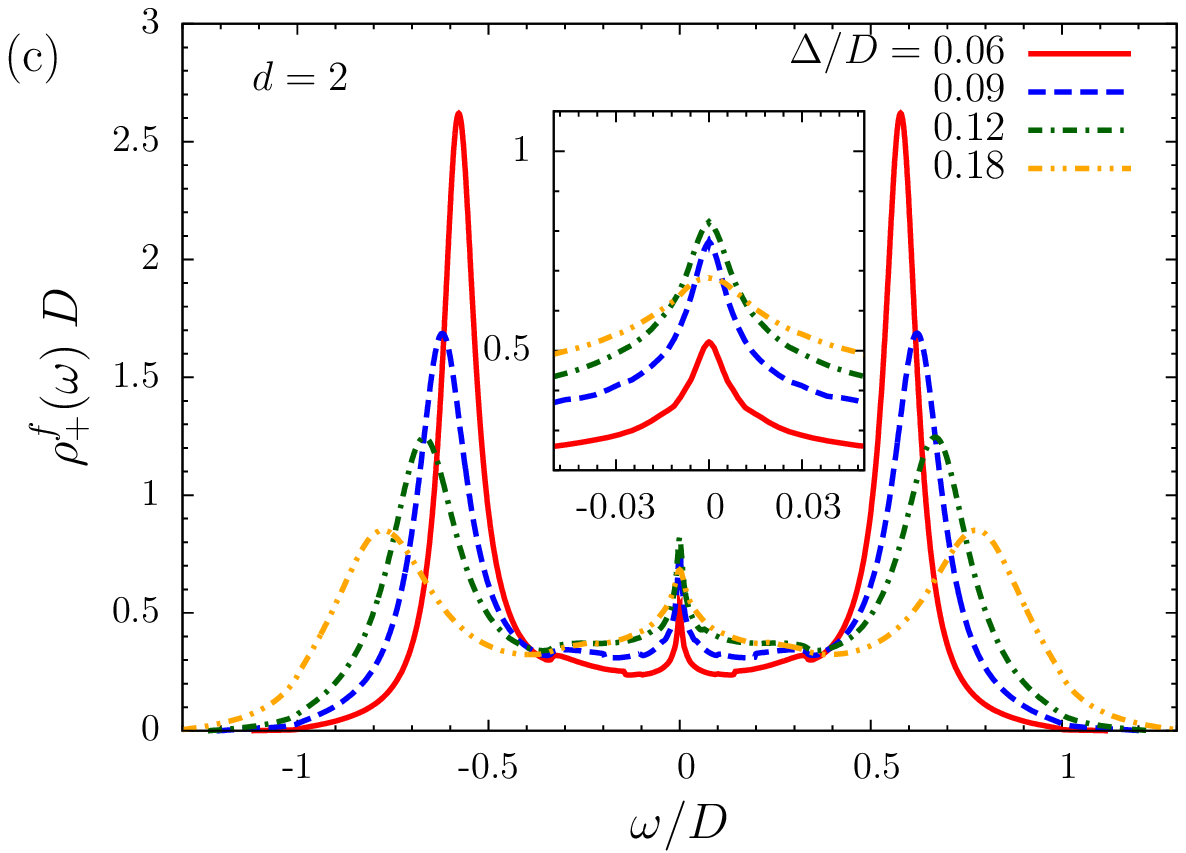}%
  \includegraphics[width=0.5\linewidth]{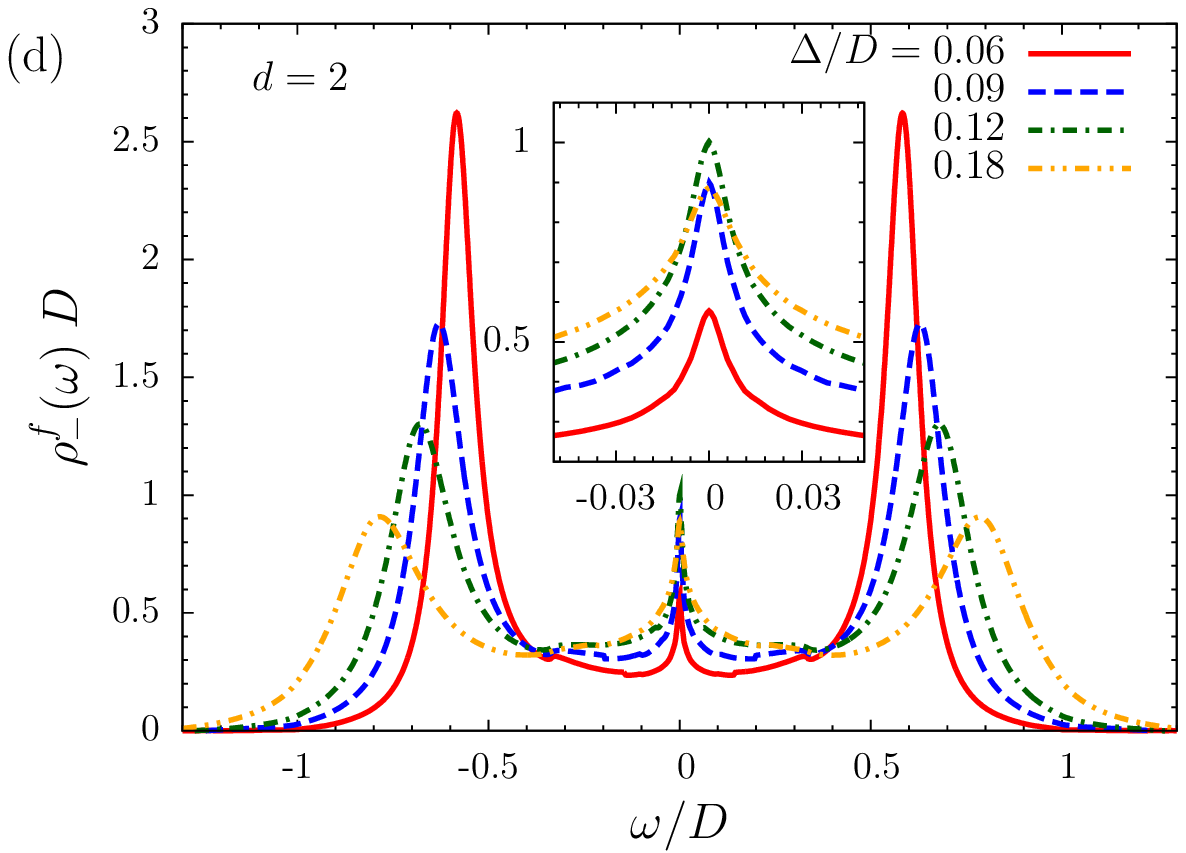}
  \caption{(Color online) Impurity spectral functions decomposed into even [(a) and (c)] and odd
    [(b) and (d)] contributions for distance   $|\vek d|=1$ [(a) and (b)]
    and $|\vek d|=2$ [(c) and (d)] for various Anderson widths.
    The inset shows a close-up of the low energy region. 
    All other parameters are as in Fig.~\ref{fig:res10}.
  }
  \label{fig:res11}
\end{figure}
The influence of molecular correlations
favoring even and odd one-particle states via an induced effective transfer
can again be derived from a decomposition of the spectra in the even
and odd parity components. These are shown in Fig.~\ref{fig:res11}
for the same parameters as in Fig.~\ref{fig:res10}.
It is obvious, that the indirect hopping is much less effective 
in producing  even-odd splitting when compared to the direct 
hopping of the previous section. The even and odd spectral functions  are
very similar to each other and only for the small distance $d=1$ 
moderate asymmetries occur.
The splitting is noticeable, though not nearly as strong as in Fig.~\ref{fig:res8}.
But interestingly, the splitting of the
low-energy quasiparticle peak at the Fermi level is much more pronounced
than the imbalance of the high-energy Hubbard peaks. This supports 
the idea that nonlocal correlations cause the coherence in the low-energy region where  
the Kondo effect develops. They also lead to pseudogap
formation and in consequence to the notion of 
a impending quasiparticle bandstructure, a viewpoint already 
adopted above.        

For  $d=2$ such an effect is barely visible and an asymmetry can not be observed 
in the spectral functions. The odd channel has a slightly larger 
spectral weight  in the many-body resonance around the Fermi level 
as compared to  the even spectra. 

\begin{figure}[ht]
  \centering
  \includegraphics[width=0.95\linewidth]{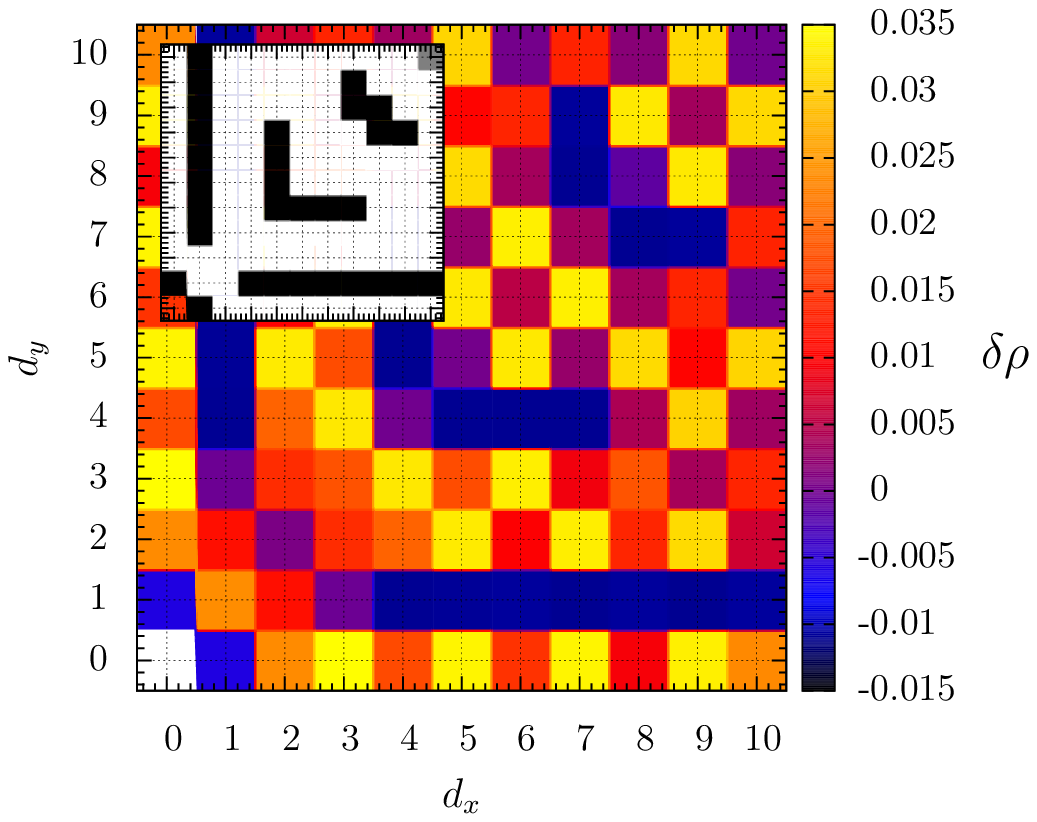}
  \includegraphics[width=0.95\linewidth]{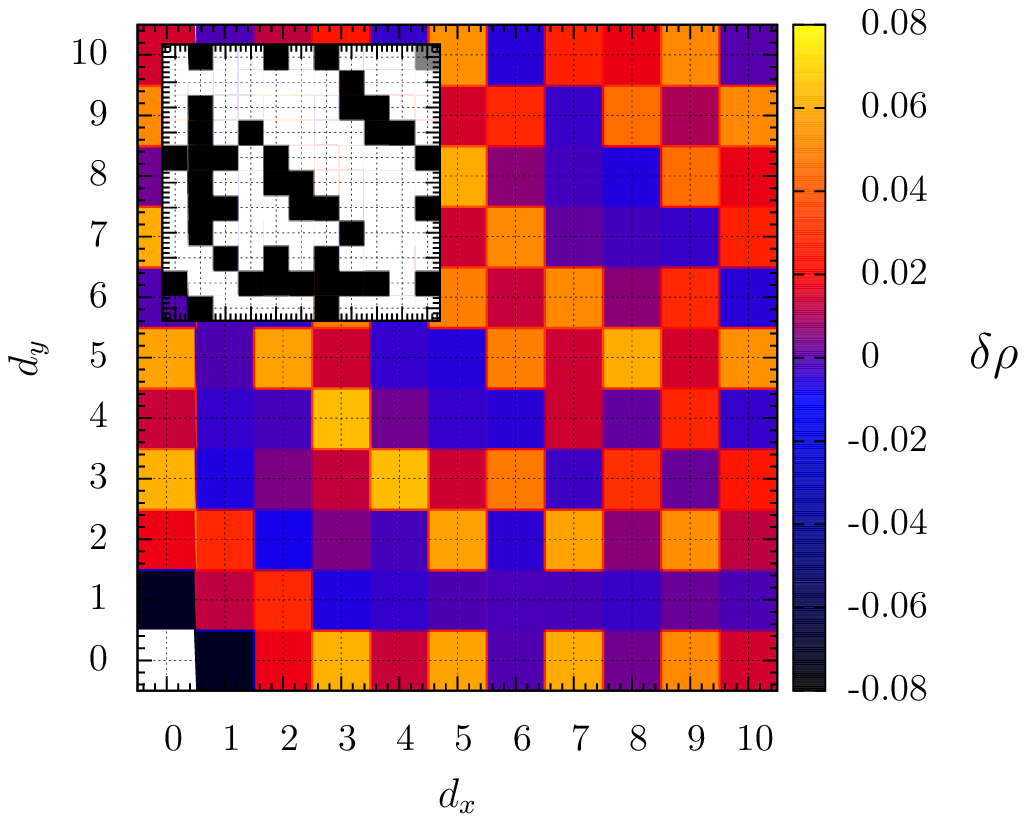}
  \caption{Height of the two-impurity spectral function at 
    $\omega=0$ relative to its value without the second impurity
    [see Eq.~\eqref{eq:dr}]. (a) $\Delta=0.11D$ 
    (b) $\Delta=0.16D$.
    One impurity is located at the origin, while the other
    at the distance vector 
    $\vek d=d_x\vek e_x+d_y\vek e_y$.
    The insets show the sign of this quantity, where black denotes
    $\Delta\rho<0$ and white  $\Delta\rho>0$.
    All other parameters are as in Fig.~\ref{fig:res10}.
  }
  \label{fig:res12}
\end{figure}

The investigation of the effective transfer- and
exchange can easily be extended to other relative positions of the
two impurities in a cubic  lattice. As one might expect,
an oscillation of physical quantities which can be associated with
ferromagnetic or antiferromagnetic tendencies is found with
varying distance. One example of such an oscillating quantity is the
height of the many-body resonance right at the Fermi level
in comparison to its  value without the second impurity, i.e.\
\begin{align}
  \label{eq:dr}
  \delta\rho= \rho(\om=0)-\rho_{|\vek d|=\infty}(\om=0).
\end{align}
We present this quantity in Fig.~\ref{fig:res12} for two different 
hybridization strengths as a function of the
two-impurity distance $\vek d=d_x\vek e_x+d_y\vek e_y$ 
in a two-dimensional plane, where one impurity is always located in the origin.
Red and yellow colors imply a larger height of the two-impurity
spectral function than in a SIAM, while  black and blue colors 
indicate a suppression of $ \rho(\om=0)$.

Regions with enhancement and suppression of the many-body resonance can be
clearly identified. While a suppression can easily be explained by the 
antiferromagnetic exchange coupling, the reason for an enhancement is not as clear.
We attribute it to the effect of the dynamic non-diagonal hybridization,
producing a constructive interference  and an effectively larger
total hybridization. 
However, the spatial pattern is not simply determined by the Manhattan distance,
i.e.\ the number of elementary hoppings between the two impurities. It appears,
that  more  complicated interferences of single-particle and interaction effects
obviously play a role.  For very large distances these short-distance effects 
should be irrelevant and an asymptotic form like Eq.~\eqref{eq:9} is expected.

\section{Conclusion}
\label{sec:conclusion}
In this work we have presented a novel two-impurity solver which is
based on direct perturbation theory with respect to 
the hybridization. It is capable of treating any kind of 
direct interaction of the two-impurities as well as the 
dynamical indirect coupling via the conduction band.

Including separately different couplings of the two impurities
 we carefully investigated the role played by 
each and the resulting physical mechanisms. 
In situations, where the impurities are not coupled 
via the conduction electrons but only directly, we found 
that a ferromagnetic exchange leads to a narrowing of the many-body 
resonance which is characteristic of a higher-spin Kondo effect. 

For antiferromagnetic coupling, the inter-impurity singlet
formation competes with two separate local Kondo effects. This 
leads to the suppression of the Kondo effect due to  
singlet-triplet splitting at small enough 
temperature which is clearly signalled by a pseudo-gap opening
in the many-body resonance at the Fermi level.  

Interestingly, such a pseudo-gap is also induced without direct exchange 
coupling but  instead including a direct single-electron hopping
between the impurities. This produces a splitting of bonding and anti-bonding
two-impurity orbitals, i.e.\ of even and odd parity components of the spectral function.
Additionally, an effective antiferromagnetic exchange is generated
by a finite hopping.
This also supports the formation of a gap in the many-body resonance
which, however,  is accompanied 
by sharp side-peaks of the even and odd excitations. 

In the generic scenario without direct interactions
between the impurities but with  couplings  generated 
indirectly via the 
conduction electrons the situation is not as clear. Signatures
of the induced RKKY interaction are observable in form of
oscillatory behavior with varying distance between the 
impurities. It seems that the induced magnetic exchange interactions
are the dominant source of competition with isolated Kondo effects.
Molecular correlations and splitting of even and odd parity channels
induced by an  effective single-particle hopping
between the impurities are only observable for smallest distances.
Even though the total size of these  effects  is rather small,
the low-energy many-body physics seems to be very sensitive to this
kind of perturbation.

Although the well known Doniach-scenario
captures essential aspects of the physics involved,
we are led to the conclusion that
the coupled two-site cluster 
exhibits  a much richer physical behavior, in particular 
at small impurity distances. For this reason  
the dynamical non-diagonal hybridization function needs 
to taken into account. 

The ability of the solver to work with arbitrary 
two-impurity distances and arbitrary hybridization functions
allows for utilization in a nonlocal extension of 
dynamical mean-field theory.\cite{pruschke:dmftNCA_HM95,*georges:dmft96} 
We propose such a scheme where two-site correlations are included
in a separate publication\cite{jabbenNonlocal12}
and the results with the two-impurity ENCA as impurity solver
are very promising.

\begin{acknowledgments}
  We thank Eberhard Jakobi for fruitful discussions,
  and the NIC, Forschungszentrum J\"ulich,  for their supercomputer support under 
  Project No.\ HDO00. 
  SS acknowledges financial support from the Deutsche
  Forschungsgemeinschaft under Grant No.\ AN 275/6-2. 
\end{acknowledgments}




\end{document}